\documentclass[amsmath,amssymb,twocolumn,amsfonts,superscriptaddress]{revtex4-2}
\usepackage{graphicx}
\usepackage{epsfig}
\usepackage{bm}
\usepackage{bbm,mathbbol}
\usepackage{braket,bigints}
\usepackage[T1]{fontenc}
\usepackage[utf8]{inputenc}
\usepackage{lmodern}
\usepackage{geometry}
\usepackage{amsmath,amssymb,amsthm,mathtools,bm}
\usepackage{physics}
\usepackage{siunitx}
\usepackage{enumitem}
\usepackage{hyperref}
\usepackage{stmaryrd,mathtools}
\usepackage{csquotes}
\usepackage{placeins}

\hypersetup{colorlinks=true,linkcolor=blue,citecolor=blue,urlcolor=blue}

\def\UoC{\small{Department of Physics, University of Crete, 70013 Heraklion, Greece}}

\def\UoR{\small{ENSSAT, Université de Rennes, 22300 Lannion Cedex, France}}

\usepackage{tikz,xcolor,hyperref}
\definecolor{lime}{HTML}{A6CE39}
\DeclareRobustCommand{\orcidicon}{%
	\begin{tikzpicture}
	\draw[lime, fill=lime] (0,0) 
	circle [radius=0.16] 
	node[white] {{\fontfamily{qag}\selectfont \tiny ID}};	\draw[white, fill=white] (-0.0625,0.095) 
	circle [radius=0.007];	\end{tikzpicture}
	\hspace{-2mm}}
\foreach \x in {A,...,Z}{%
	\expandafter\xdef\csname orcid\x\endcsname{\noexpand\href{https://orcid.org/\csname orcidauthor\x\endcsname}{\noexpand\orcidicon}}
}

\newcommand*{\rom}[1]

\begin{document}

\title{Constraining  axion-like dark matter with a radio-frequency atomic magnetometer}

\author{A. Rigoulet}
\thanks{These authors contributed equally to this work.}
\affiliation{\UoC}
\affiliation{\UoR}

\author{S. Nanos\orcidB{}}
\thanks{These authors contributed equally to this work.}
\affiliation{\UoC}

\author{I.\,K. Kominis}
\affiliation{\UoC}
\affiliation{School of Science, Zhejiang University of Science and Technology, Hangzhou 310023, China}

\author{D. Antypas\orcidD{}}
\email{dantypas@physics.uoc.gr}
\affiliation{\UoC}

\date{\today}

\begin{abstract}
We report on a broadband search for axion-like-particle (ALP) interactions using a radio-frequency-operated $^{87}\mathrm{Rb}$ atomic magnetometer. The instrument provides wide spectral coverage and sensitivity to an oscillating  pseudomagnetic field that may be generated by the gradient coupling of the ALP field to the constituent fermions of atoms. We search for an ALP-gradient signature in the mass range $2.40\times10^{-10}\,\mathrm{eV}/c^{2}$--$2.11\times10^{-9}\,\mathrm{eV}/c^{2}$. No statistically significant signatures of an oscillating magnetic field are observed, and we derive upper limits on the corresponding ALP-proton, -neutron and -electron couplings, $g_{\alpha pp}$, $g_{\alpha nn}$ and $g_{\alpha ee}$, respectively. The result on $g_{\alpha pp}$ improves over previous laboratory searches, while the limits on $g_{\alpha nn}$ and $g_{\alpha ee}$ complement earlier laboratory searches and astrophysical bounds. The work extends searches for ALP-fermion interactions into a mass region largely unexplored in a dark-matter context, demonstrating the potential of our method for broadband axion-like particle searches targeting the Galactic dark-matter halo. 
\end{abstract}

\maketitle

\section{Introduction}
There is direct astronomical evidence for the existence of dark matter (DM) \cite{KimballBook}, a non-luminous form of matter assumed to make up the majority of mass in the Universe\,\cite{NavasPRD2024}. In spite of numerous terrestrial searches, its nature and properties remain unknown.
Within a motivated class of scenarios \cite{BertoneNature2018}, DM consists of spin-0 particles with mass in the range $10^{-22}$\,eV/c$^2-10$\,eV/c$^2$ \cite{KimballBook}.  
Such ultralight dark matter (UDM) particles are expected to form a classical field that oscillates at a frequency close to the underlying Compton frequency. 
A prominent candidate is the so-called QCD axion \cite{PreskillPLB1983, AbbottPLB1983, DinePLB1983, MarshPhysRep2016}, a pseudoscalar particle originally introduced to explain the observed suppression of CP violation in the strong force. Precision experiments \cite{GrahamARNPC2015, AntypasSnowmass2022} can search for interactions between axion-like particles and Standard Model particles.

Here we report on a search for axion-like particle (ALP) interactions with fermions, using a radio-frequency atomic magnetometer \,\cite{SavukovPRL2005, SavukovJMR2007}. By probing coupling to atomic spins we constrain ALP interactions with the proton, neutron and electron  in the ALP mass range $2.40\times 10^{-10}$\,eV/$c^2 -2.11\times 10^{-9}$\,eV/$c^2$.  Owing to the extended experimental bandwidth of the employed detector, we are able to  carry out our search in a broad mass range that has  remained unexplored in the majority of searches with spin-based probes, which are mostly sensitive  at lower masses \cite{WUPRL2019,GarconSciAdv2019,LeePRX2023,GMNatComm2025, WeiRPP2025}. The ALP-UDM scenario investigated here is distinct from fifth-force and astrophysical constraints, which pertain to non-dark-matter scenarios \cite{IrastorzaSD2018}.

This paper is organized as follows: in Sec.~\ref{sec:phenomenology} we summarize the phenomenology of ALP-gradient couplings to fermions and their mapping to an experimental observable. In Sec.~\ref{sec:experiment} we describe our detection approach and the radio-frequency magnetometer employing a spin-polarized $^{87}$Rb vapor. In Secs.~\ref{sec:dataacquis}~and~\ref{sec:dataaanal} we discuss the experimental data acquisition and data analysis, respectively. In Sec.~\ref{sec:translating} we translate measured spin observables into fundamental ALP-fermion couplings, the measured constraints on which are presented in \ref{sec:constraints}. We conclude in Sec.~\ref{sec:conclusions} with a summary and outlook.  Supporting information is provided in the Appendices.

%

\section{Phenomenology of ALP-gradient coupling to fermions\label{sec:phenomenology}} 
The collective field of UDM particles can be approximately  expressed as:
\begin{equation}
a(\mathbf{r}, t) = a_0 \cos(2\pi \nu_a t - \mathbf{k} \cdot \mathbf{r} + \phi),
\label{eq:alp_field}
\end{equation}

\noindent where $\alpha_0$ is the field amplitude, $\mathbf{k} = m_a \mathbf{v}_a/\hbar$ is the wave vector corresponding to a particle with mass  $m_\alpha$ and velocity $\mathbf{v}_\alpha$, $\phi$ is a random phase, $2\pi\hbar$ is the Planck constant and $c$ is the speed of light. The field oscillates at the Compton frequency $\nu_a = m_a c^2/2\pi \hbar$, and its amplitude is set by the local 
dark matter density: $\alpha_{0}=\hbar\sqrt{2\rho_{\rm DM}}/(m_{\alpha}c)$, with $\rho_{\rm DM}\approx 0.3$\,GeV/cm$^3$ \cite{WechslerAnnRevAstro2018}.  

A more accurate description of the UDM field is obtained  by considering the velocity dispersion of the particles in the Galactic DM halo, about its mean value $\mathit{v}_0=$220\,km/s. For an observer on Earth, this dispersion results in a spread of UDM particle frequencies; a particle with velocity $v$ will appear having frequency close to $(1+v^2/2c^2)\nu_\alpha$.   Moreover, the phases $\phi$ of different ALP oscillations  are assumed to be uncorrelated \cite{FosterPRD2018}. As a result, the field is best described as a superposition of a large number of oscillators with slightly different frequencies and random phases, leading to random fluctuations of the amplitude $\alpha (\mathbf{r},t)$, having a finite coherence time $\tau_a \approx \left( \nu_a v_a^2/{c^2} \right)^{-1}$. 

ALP coupling to fermions is described by the Hamiltonian
\begin{equation}
H_{\alpha f} = g_{\alpha ff}\nabla \alpha \cdot \mathbf{ S}_f,
\label{eq:IntHam}
\end{equation}
\noindent where $\mathbf{ S}_f$ is the fermion spin (\textit{f=e,p,n}), $g_{\alpha ff}$ is the ALP coupling to the fermion, and  
$\nabla \alpha$ is the ALP gradient, approximately given by $\nabla \alpha \approx (a_0 m_a /\hbar)\mathbf{v}_a$. This gradient coupling is analogous to the Zeeman interaction $H=\hbar\gamma_f \mathbf{S}_f\cdot \mathbf{ B}$,  with 
 an effective (pseudomagnetic) field given by $B_{\alpha}=g_{\alpha ff}\nabla\alpha/\gamma_f$, where $\gamma_f$ is the fermion's gyromagnetic ratio. 
 
Interactions of the form given in Eq.\,(\ref{eq:IntHam}) between an axion-like field and the fermionic constituents of an atom give rise, at low energies, to an effective torque on the atomic spin degrees of freedom.
The resulting atomic dynamics can be written as an effective precession of the total atomic angular momentum $\mathbf F=\mathbf S+\mathbf I$, i.e. $d\mathbf F/dt=\boldsymbol{\Omega}_{\alpha}\times \mathbf F$,
where $\boldsymbol{\Omega}_{\alpha}$ is the axion-induced precession frequency determined by the fundamental axion-fermion coupling and by atomic-structure factors. For comparison with magnetometer sensitivities, which are conventionally expressed in terms of magnetic fields, it is convenient to define an
effective axion-induced pseudo-magnetic field $\mathbf B_{\alpha}$ via $\boldsymbol{\Omega}_{\alpha}=\gamma_F\,\mathbf B_{\alpha}$, where $\gamma_F=\gamma_e/(2I+1)$ is the gyromagnetic ratio of the hyperfine state of alkali atoms with nuclear spin $I$ in the electronic ground state,
and $\gamma_e=2\pi\times 28~\mathrm{GHz/T}$ is the free-electron gyromagnetic ratio. The effective axion-induced pseudo-magnetic field is then given by
\begin{equation}
\mathbf{B}_{\alpha}
=
\frac{\chi_f\, g_{\alpha ff}}{\gamma_F}\,\nabla\alpha .
\end{equation}
Here $\chi_f$ is a dimensionless conversion factor that accounts for the projection of the fermion spin operator $\mathbf{S}_f$ onto the atomic angular momentum probed by the magnetometer, thereby relating the fundamental axion-fermion coupling to the measured atomic response
\cite{KimballNJP2015, StadnikEPJC2015}, as will be detailed in Sec. \ref{sec:translating}.

For an ultralight axion field constituting the local dark matter density $\rho_{\mathrm{DM}}$, the axion field gradient oscillates at the Compton frequency $\nu_a$, and the pseudo-magnetic field will read
\begin{equation}
\mathbf{B}_{\alpha}(t) =\frac{\chi_f\, g_{\alpha ff}}{\gamma_F}
\sqrt{2\hbar c\,\rho_{\mathrm{DM}}}\,
\sin(2\pi \nu_a t)\,\mathbf{v}_a ,
\label{eq:pseudofield}
\end{equation}
where $\mathbf{v}_a$ is the axion velocity in the laboratory frame. This field oscillates at the ALP frequency $\nu_\alpha $ and points along the direction of the ALP velocity $\mathbf{v}_a$. 
%
\section{Experiment\label{sec:experiment}}
Magnetic resonance techniques \cite{BudkerPRX2013,WalterPRD2025, Bloch:2021vnn, GMNatComm2025, LeePRX2023, WUPRL2019, WeiRPP2025} can facilitate the search for an ALP-gradient interaction embodied by Eq.\,(\ref{eq:IntHam}). Here we employ a radio-frequency magnetometer \cite{SavukovPRL2005,SavukovJMR2007} working with a vapor of spin-polarized $^{87}$Rb atoms to search for the pseudomagnetic field of Eq.\,(\ref{eq:pseudofield}). This approach allows a broadband coverage for the frequency $\nu_a$, which in this work is in the range 58$-$510 kHz.


Atomic spins are optically polarized along a leading magnetic field $\mathbf{B}_0$, which induces spin precession at the Larmor frequency $\nu_L$, set by $\mathbf{B}_0$. A small transverse magnetic field $\mathbf{B}_{\alpha\perp}(t)$ oscillating near $\nu_L$ coherently drives the spin system, generating a resonantly enhanced transverse spin component. This component, perpendicular to both $\mathbf{B}_0$ and $\mathbf{B}_{\alpha\perp}(t)$, is detected via optical Faraday rotation \cite{Demtroder}, allowing to measure $\lvert \mathbf{B}_{\alpha\perp}(t) \rvert$. Scanning $B_0$ tunes $\nu_L$, enabling sensitivity over the aforementioned frequency range.

\begin{figure*}[t]
    \centering    \includegraphics[width=0.7\textwidth]{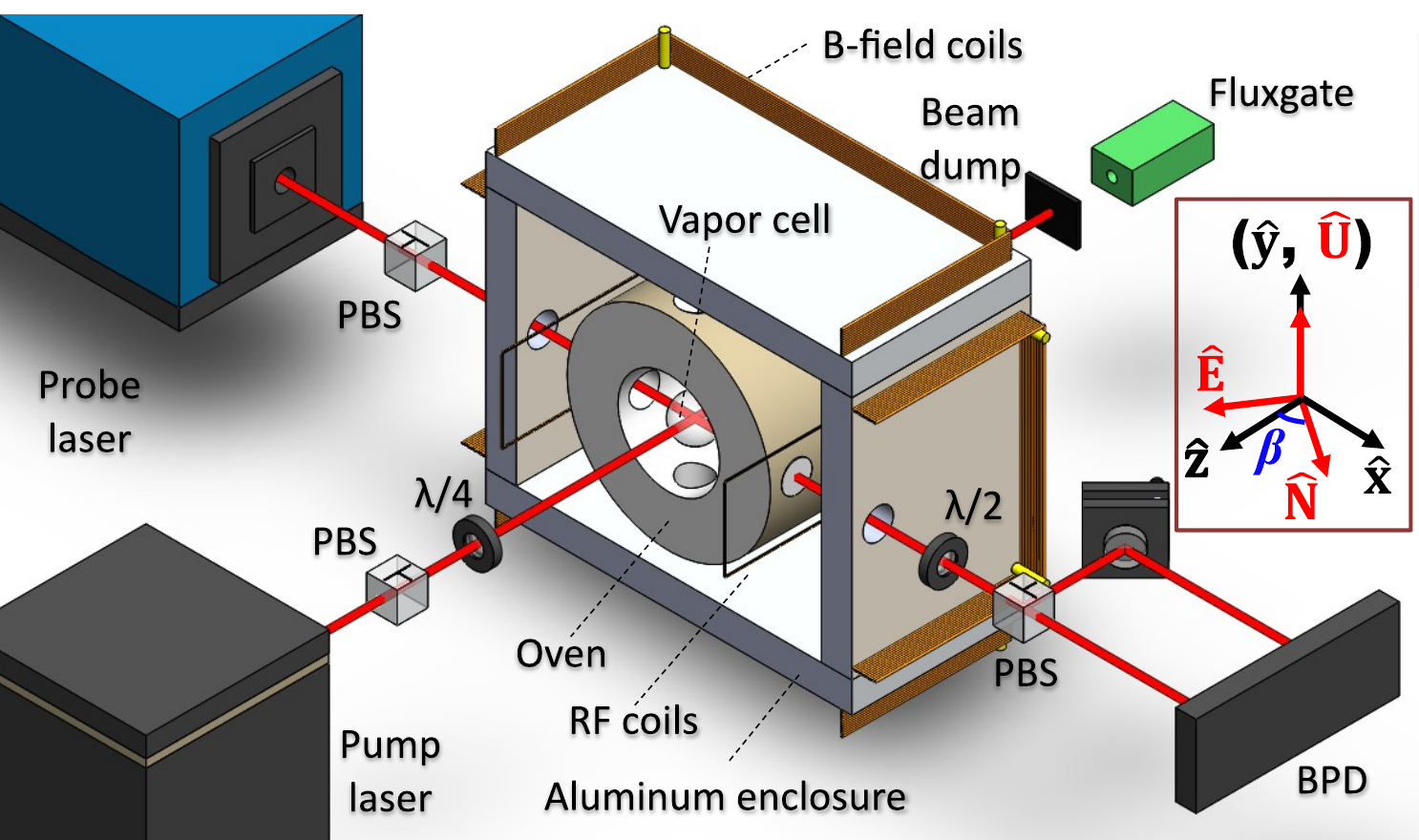}
 \caption{ 
 \small{ Simplified apparatus schematic. Rb atoms in a heated vapor cell are spin-polarized and probed using crossed laser beams. The cell is housed in an aluminum enclosure, surrounded by sets of coils to null the ambient field along $\hat{x}$ and $\hat{y}$, and set the leading field along $\hat{z}$. A set of rf coils is used to apply a known field to check the magnetic resonance response. The lab coordinate system $\hat{x}\hat{y}\hat{z}$ is shown in relation to the local coordinate system with axes $\hat{N}$ (north), $\hat{E}$ (east) and $\hat{U}$ (local zenith). Abbreviations: $\lambda/2$: half-wave plate; $\lambda/4$: quarter-wave plate; PBS: polarizing beam splitter; BPD: balanced photodetector. %
 }
}
    \label{fig:apparatus}
\end{figure*}

The atomic vapor is contained in a spherical glass cell of diameter 17.5 mm, heated at 130 \,$^\circ$C\,(Fig.\,\ref{fig:apparatus}). Atoms are spin-polarized using a circularly polarized laser beam, tuned to excite the $5S_{1/2}\rightarrow 5P_{3/2}$ optical transition at \mbox{780.2 nm}. This beam is along the applied magnetic field $\mathbf{B_0}$ pointing along $\hat{z}$. Faraday rotation induced by a field transverse to $\mathbf{B_0}$ is measured via polarimetry of a linearly polarized beam that propagates through the atomic medium along $\hat{x}$, and is tuned  $\approx 100$\,GHz below the $5S_{1/2}\rightarrow 5P_{3/2}$ transition frequency. 
The measured Rb density is $\approx 1.2\times 10^{19}$\,m$^{-3}$. The presence of $^4$He (600 torr) and N$_2$ (50 torr) serves to suppress spin decoherence 
and to improve spin polarization efficiency, respectively, but broadens the optical transition to about \,11\,GHz. This is greater than the $\approx6.83$\,GHz spacing between the 5S$_{1/2}$ $F=1$ and $F=2$ hyperfine levels. Consequently, 
the two levels are not resolved in the optical spectrum, and atoms in both hyperfine manifolds are simultaneously addressed and contribute to the magnetometric signal.

Sets of rf coils around the vapor cell are employed to apply a known ac magnetic field along $\hat{x}$ or $\hat{y}$ in order to record the magnetic resonance spectrum and determine the magnetometer sensitivity. To null the ambient magnetic field, use of a high permeability ($\mu-$metal) shield is avoided, as this is expected to partially shield the exotic field searched for \cite{KimballPRD2016}. Instead, the cell is housed in a 20-mm thick aluminum enclosure, which provides no shielding against the ambient DC magnetic field, but is effective in the \mbox{58$-$510 kHz} frequency range of our UDM search. Cancellation of the ambient field along  $\hat{x}$ and $\hat{y}$ is done with sets of coils surrounding the enclosure, while  another set is used to control the amplitude  ${B_0}$. 

Measurements of the oscillating Faraday rotation
are done through lock-in detection of the polarime-
ter output. To calibrate the apparatus sensitivity,
the magnetic resonance is recorded at a given am-
plitude $B_0$ by stepping the frequency of an applied
rf field of known amplitude around the respective
Larmor frequency. Comparison of the recorded resonance amplitude with the background noise level yields a sensitivity of   $\approx15$\,fT/$\sqrt{\rm Hz}$ at the high end of the investigated frequency range. The resonance has a linewidth (Full Width at Half Maximum-FWHM) of $\approx 4.5$\,kHz. This is far broader than the expected ALP linewidth $\delta\nu_a\approx1/\tau_{\alpha} \approx \nu_a v_a^2/{c^2}$, or 58$-$510\, mHz within the frequency range investigated \footnote{Here we follow the usual assumption in the literature that $v_{\alpha}/c\approx10^{-3}$.}. Thus, an observed ALP signal in the spectrum will acquire the magnetometer linewidth.

Since the vapor cell enclosure provides no shielding against static magnetic fields, the value of ${B_0}$ at the location of atoms includes not only the field created by the $\hat{z}$-set of coils, but also the contribution from the ambient field along  $\hat{z}$. Therefore, active stabilization of the field along $\hat{z}$ in the vicinity of atoms is implemented, to compensate for ambient drifts. This way the Larmor frequency can be set and maintained accurately, independently of ambient drifts (see Sec.~\ref{sec:Apparatus} in the Appendix).

The main systematic uncertainties in the apparatus arise from a slight imperfection in setting the Larmor frequency, as well as from imperfect knowledge of the calibrated rf fields used to convert the measured voltage response into a magnetic field response. Both are estimated to contribute at the $\sim$10\% level, resulting in an overall $\mathcal{O}(10\%)$ uncertainty in the determination of the apparatus magnetic field response.

\begin{figure*}[t]
\includegraphics[width=\textwidth]{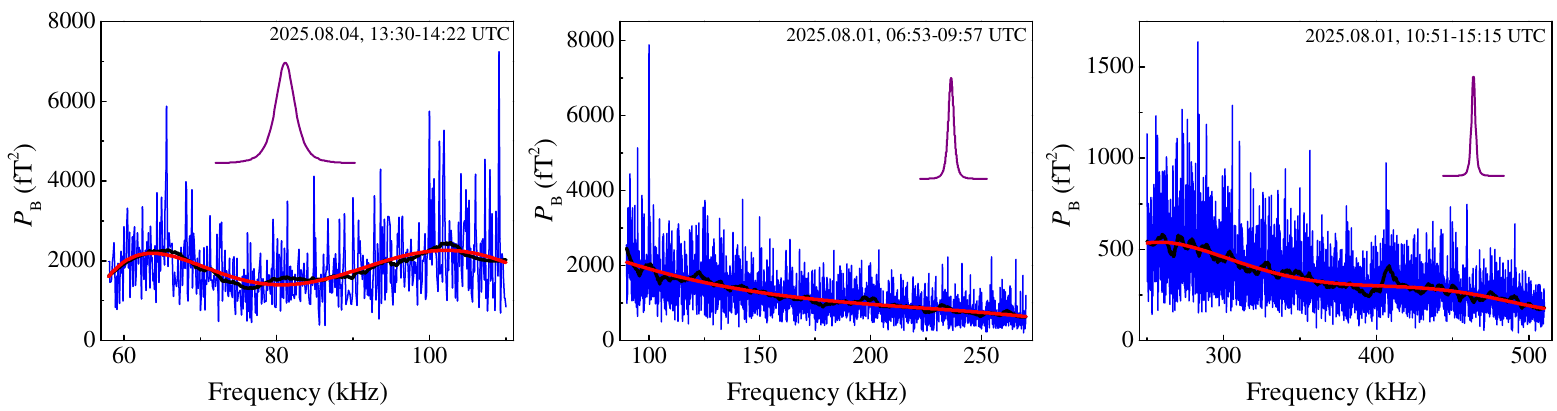}
\caption{Magnetic field power spectra (blue) acquired in the three main experimental runs.  Polynomial fits (red) to respective SG filter results (black) are used to determine the local mean noise level. The simulated lineshapes in the plots  (to scale in frequency) show the squared-Lorentzian lineshape of a magnetometer resonance.}
\label{fig:B2spectrum}
\end{figure*}
 
\section{Data acquisition\label{sec:dataacquis}}
We carry out a search for the field $\vert\mathbf{B}_{\alpha \perp} (t)\vert$ by acquiring spectra of the magnetometer response in three Larmor-frequency intervals: 58$-$110 kHz, 90$-$270 kHz, and 250$-$510 kHz. The partial frequency overlap in these experimental runs is incorporated in data analysis, as we discuss below. In each run, the leading field $B_0$ is stepped by 100 Hz in terms of $\nu_L$ and the magnitude of the lock-in amplifier voltage output is recorded at the respective frequency. The integration time per step is 4.5~s, with a 1.5~s of settling time between steps, amounting to a total of $\approx7.5$~h of data taking. 

Both before and after a run, the apparatus response is calibrated at several Larmor frequencies within the respective frequency interval, by applying a small test field and recording the magnetic resonance amplitude. The measured voltage noise spectra from the three runs are converted into magnetic field power spectra $P_B(\nu)=B^2(\nu)$, which exhibit pronounced spectral peaks at harmonics of the vapor cell heater assembly (operating at 50\,kHz), due to apparatus pickup\,\footnote{These peaks always have linewidth smaller than 500 Hz, i.e. much less than the $\approx 4.5$\,kHz width of a magnetic resonance.}. Additional measurements are acquired over narrow ranges ($\approx1$\,kHz wide) around these harmonic frequencies with the oven heater briefly powered off, and used to replace the respective data points in the power spectra. The resulting spectra are further analyzed for signatures of an ALP-gradient coupling.
\section{Data analysis\label{sec:dataaanal}}
We search for features in the three $P_B^{(k)}(\nu)$ spectra ($k=1,2,3$) that rise above the background noise level and exhibit the expected resonance linewidth (Fig.\,\ref{fig:B2spectrum}). Because the mean noise level in the data is not consistent across the recorded frequency spans, primarily due to technical noise that scales approximately as $1/\nu$, it is advantageous to normalize $P_B(\nu)$ to the mean noise level $N(\nu)$. This enables a uniform statistical analysis over the full recorded bandwidth.

The quantity $N(\nu)$ is obtained in two steps. First, a Savitzky-Golay (SG) filter with a 5-kHz window and a 2$^{\rm nd}$-order polynomial is applied to each $P_B(\nu)$ spectrum. While the filter captures the local mean noise level, it can also partially track narrow peaks of width comparable to the 
magnetometer linewidth. Therefore, in a second step, a $5^{\rm th}$-order polynomial is fitted to the SG-filter output. This fit is taken as the mean noise level $N(\nu)$; it varies slowly with $\nu$, effectively eliminating the risk of reproducing spectral peaks and thereby preventing distortion of the normalization. This two-step approach yields better performance than a direct polynomial fit to the spectra, particularly near the edges of the three spectra.

The resulting normalized spectra \mbox{$P^{(k)}_n(\nu)=P_B^{(k)}(\nu)/N^{(k)}(\nu)$} are merged into a single spectrum for further analysis. In the regions of spectral overlap (i.e. within either 90$-$110\,kHz or 250$-$270\,kHz) the spectra are combined with tapering using the weight coefficients of a Hann window centered at $\nu_0=$ 100\,kHz or 260\,kHz. We compute the normalized spectrum within these overlap regions as:
%
%
%
\begin{widetext}
\begin{align}
P_n(\nu) =
\begin{cases}
w^{(1)}(\nu)P^{(1)}_n(\nu)+w^{(2)}(\nu)P^{(2)}_n(\nu), & 90 \le \nu \le 110~\mathrm{kHz}, \\
w^{(3)}(\nu)P^{(2)}_n(\nu)+w^{(4)}(\nu)P^{(3)}_n(\nu), & 250 \le \nu \le 270~\mathrm{kHz},
\end{cases}
\label{eq:Ptapering}
\end{align}
\end{widetext}

\noindent where $w^{(1)}(\nu)=\rm{cos}^2(\pi \mathit{x}^{(1)}(\nu)/2)$,   $w^{(2)}(\nu)=1- w^{(1)}(\nu)$, $w^{(3)}(\nu)=\rm{cos}^2(\pi \mathit{x}^{(3)}(\nu)/2)$,   $w^{(4)}(\nu)=1- w^{(3)}(\nu)$, with $\mathit{x}^{(1)}(\nu)=(\nu-100\,\textrm{kHz})/( \rm 20\, kHz)$,  and $\mathit{x}^{(3)}(\nu)=(\nu-260\,\textrm{kHz})/( \rm 20\, kHz)$.
~The merged spectrum $P_n(\nu)$ has unit mean power and is well approximated by a Gamma probability distribution function (PDF) with real-valued shape and scale parameters (see Section \ref{sec:DetThres} in the Appendix).\\

To identify a magnetic field signal in $P_n(\nu)$ with optimal sensitivity, we apply a matched filter to the data, similarly as in \cite{WalterPRD2025, BrubakerPRD2018}. This filter is evaluated within a $\pm10$\,kHz window, centered on each frequency data point. Because the analysis is performed in power, the signal lineshape acquires the squared-Lorentzian response of the magnetometer. Thus, the filter kernel is the function $L^2(\nu)$, where $L(\nu)$ is a Lorentzian of unity peak amplitude and FWHM of 4.5 kHz, i.e. the expected resonance linewidth. The output of this filter, $P^f_n(\nu)$ is examined for statistically significant peaks above the noise background.

The sensitivity of our experiment is evaluated through a hypothesis test, namely to determine whether any feature in the filtered spectrum, $P^f_n(\nu)$, exceeds what is expected from noise at a specified confidence level. To establish a statistically valid detection threshold, we perform $10^5$ Monte Carlo realizations of synthetic noise spectra, each having the same statistical properties as the experimental spectrum $P_n(\nu)$. Every realization is processed through the matched filter to produce simulated filtered spectra $P^f_n(\nu)$, and the maximum value in each, $X_{\mathrm{max}}$ , is recorded. From the ensemble of maxima, we construct the cumulative distribution function, $F_{X_{\mathrm{max}}}(x)$, of $X_{\mathrm{max}}$. The 95$^{\rm th}$ percentile of this distribution, $X_{\mathrm{thr}}$, is determined by the probability
\[
\mathbb{P}(X_{\mathrm{max}} \le X_{\mathrm{thr}}) = F_{X_{\mathrm{max}}}(X_{\mathrm{thr}}) = 0.95.
\]
This threshold defines the hypothesis test: any frequency bin $\nu$ in the experimental spectrum for which
\mbox{$P^f_n(\nu) > X_{\mathrm{thr}}$} has only a 5\% probability to arise from random noise, and is considered a candidate peak and flagged for further experimental investigation.
In our analysis, the look-elsewhere effect\,\cite{Scargle1982} associated with scanning multiple independent frequency points is accounted for by constructing the detection threshold from Monte Carlo realizations of the full frequency range probed (58-510\,kHz). Since for each realization we evaluate the maximum fluctuation $X_{\mathrm{max}}$ across the full  spectrum, the resulting threshold has a global significance.

\begin{figure}[t]
    \centering    \includegraphics[width=\columnwidth]{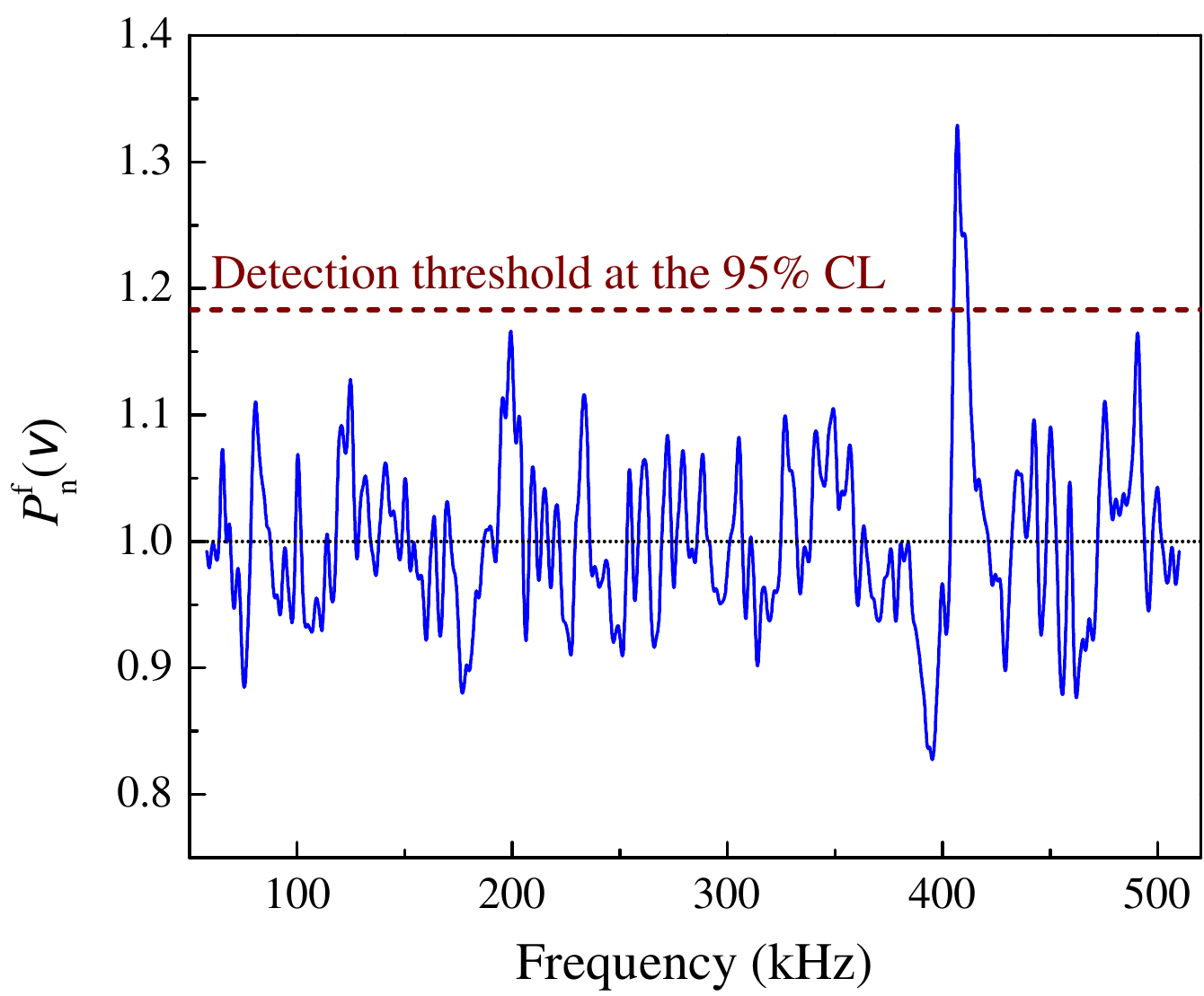}
 \caption{ 
 \small{Matched filter output $P_n^f(\nu)$ of the main data set $P_n(\nu)$ and detection threshold $X_{\rm thr}=1.183$ at the 95\% confidence level. }
}
    \label{fig:FilteredB2}
\end{figure}

We show the experimental spectrum $P^f_n(\nu)$ in Fig.\,\ref{fig:FilteredB2}, along with the computed detection threshold $X_{\rm thr}=1.183$ at the 95\% confidence level. We note that a single peak at $\approx\,$407\,kHz exceeds the $X_{\rm thr}$, and thus requires further investigation. Using additional data sets in the same frequency range, we determine that this peak is not reproducible (see Section\,\ref{sec:Outliers} in the Appendix). Within reasonable confidence, we can therefore exclude the peak from being an ALP-induced field. We assign the observed signal to intermittent amplitude noise of the optical pumping laser. At the level of sensitivity of the current work, we find no other significant outliers in the data.

To convert the detection threshold into a quantitative limit on potential signals, we perform an additional set of Monte Carlo simulations in which synthetic spectra include both noise sampled from the same PDF as that of the experimental spectrum $P_n(\nu)$, and an injected Lorentzian-squared signal with amplitude $A^2$. The Lorentzian itself has a FWHM matching the magnetometer linewidth. At many different frequencies $\nu$, we determine the amplitude $A_{\rm thr}^2(\nu)$ required for the filtered spectrum to reach the threshold $X_{\mathrm{\rm thr}}$.  This  quantity can then be directly translated into an upper limit on the corresponding magnetic field. Away from spectral edges the computed threshold  $A_{\rm thr}^2(\nu)$ is uniform, with $A_{\rm thr}^2(\nu)$=0.443. Near the edges of the recorded bandwidth, however (to within 2\,kHz from either 58\,kHz or 510\,kHz), it rises, up to 0.68, reflecting  the reduced sensitivity near the spectral boundaries. This effect is taken into account when reporting limits near the spectral edges. 

A limit on the effective ALP-induced field can be calculated using the determined value for $A_{\rm thr}^2(\nu)$:
\begin{equation}  \big|B_{\alpha\perp}(\nu) \big]^{(k)}_{\rm lim} = \sqrt{N^{(k)}(\nu)\big[A^2_{\rm thr}(\nu)\big]^{(k)}},
\label{eq:Balimit}
\end{equation}
%

%
\noindent with $k=1,2,3$ labeling the respective experimental runs of Fig.\,\ref{fig:B2spectrum}.
The established 95\% confidence-level limit on the field amplitude varies from $\approx 30$~fT at low frequencies to $\approx 10$~fT at high frequencies. This is consistent with the measured noise floors of $\approx 50$~fT$/\sqrt{\rm Hz}$ and $\approx 15$~fT$/\sqrt{\rm Hz}$, respectively, taking into account the incoherent, power-based ($B^2$) analysis and the effective integration time of $\approx 200$~s per frequency, given the 4.5~kHz magnetic-resonance linewidth.


\section{TRANSLATING ATOMIC SPIN OBSERVABLES INTO ALP COUPLINGS\label{sec:translating}}
Here we derive the conversion factor $\chi_f$ [see\,Eq.\,(\ref{eq:pseudofield})] required to extract the couplings of the ALP to the electronic or nucleon spins from the measured atomic spin observables\,\cite{KimballNJP2015}, which involve the total atomic spin $\mathbf{F}$. Indeed, consider an alkali atom with electronic spin $\mathbf{S}$, nuclear spin $\mathbf{I}$, and total spin $\mathbf{F}=\mathbf{S}+\mathbf{I}$. The ground state is split into the upper and lower hyperfine manifolds, with $F=a\equiv I+\tfrac12$ and $F=b\equiv I-\tfrac12$, respectively. Suppose that the atoms are optically pumped into a spin-temperature state\, \cite{AppeltPRA1998} along the $z$-axis, and probed along the $x$-axis. The probe laser probes the atomic spin observable $g_aF_{a,x}-g_bF_{b,x}$\, \cite{MouloudakisPRA2023}, where $F_{a,x}$ and $F_{b,x}$ are the $x$-components of the total spin in the upper and lower manifold, respectively. For a probe laser detuning much larger than the optical linewidth, as is the case in this experiment, the coupling constants $g_a$ and $g_b$ are roughly equal\, \cite{MouloudakisPRA2023}, thus the probed observable is $O_x=F_{a,x}-F_{b,x}$. However, the ALP couples directly to the electron spin $\mathbf{S}$, and/or the proton spin $\mathbf{S}_p$, and/or the neutron spin $\mathbf{S}_n$. We need to find how those couplings relate to the measured observable $O_x$.
\subsection{Coupling to the electronic spin}
In a fixed-$F$ hyperfine manifold, any rank-1 vector operator must be proportional to $\mathbf F$, so \mbox{$\mathbf S|_F=c_F\,\mathbf F|_F$}. Taking the scalar product with $\mathbf F$ gives $\mathbf S\cdot\mathbf F=c_F F(F+1)$, where we set $\hbar=1$. Using $\mathbf F^2=\mathbf S^2+\mathbf I^2+2\mathbf S\cdot\mathbf I$ we find
\begin{equation*}
\begin{aligned}
\mathbf S\cdot\mathbf F= &
\mathbf S^2+\mathbf S\cdot\mathbf I \\
=&\mathbf S^2+\tfrac12(\mathbf F^2-\mathbf S^2-\mathbf I^2)\\
=&\tfrac12[F(F+1)+S(S+1)-I(I+1)],
\end{aligned}
\end{equation*}
\noindent hence:
\begin{equation*}
c_F=[F(F+1)+S(S+1)-I(I+1)]/2F(F+1). 
\end{equation*}
With $S=1/2$ we have $c_a=1/(2I+1)$ and $c_b=-1/(2I+1)$. Therefore, in the upper manifold $\mathbf S=\mathbf F_a/(2I+1)$ and in the lower manifold $\mathbf S=-\mathbf F_b/(2I+1)$. Neglecting hyperfine coherences (oscillating at the hyperfine splitting), we combine these into the block-diagonal operator identity $\mathbf S\simeq(\mathbf F_a-\mathbf F_b)/(2I+1)$. Thus the electron-ALP coupling Hamiltonian of the form $H_{ae}\propto\mathbf{S}\cdot\mathbf{B}_a$ can be written as $H_{ae}\propto\chi_e(\mathbf{F}_a-\mathbf{F}_b)\cdot\mathbf{B}_a$, where $\chi_e=1/(2I+1)$. In light of Eq.~\eqref{eq:pseudofield}, this shows that the atomic spin observable directly transduces the axion-electron coupling, as the factor $\chi_e$ exactly compensates the appearance of $\gamma_F$ in the effective pseudomagnetic field.
 \subsection{Coupling to the nucleon spin}
The case of axion-nucleon coupling is slightly more complicated, since the nuclear spin $\mathbf{I}$ is not (effectively) proportional to the measured observable $\mathbf{O}$, as is $\mathbf{S}$. We first provide the conversion factor from the nuclear spin to the atomic observable $\mathbf{O}$, and then from the nucleon spin to the nuclear spin. We note that for zero axion-nuclear coupling, the atoms will remain in their initial state optically pumped along $\hat{z}$, for which $\langle I_x\rangle=\langle O_x\rangle=0$. A non-zero coupling, given by the Hamiltonian $H_{\rm int}=\chi I_y$, will slightly tilt $\langle I_z\rangle$, and hence induce a change $\delta \langle I_x\rangle$ away from zero, which will further translate into a change $\delta \langle O_x\rangle$. Indeed, within the short time interval $\delta t$,  the interaction $H_{\rm int}$ generates a small rotation by an angle ($\hbar=1$) $\theta=\chi\delta t$ about the $y$-axis in the spin space defined by $\boldsymbol{I}$. Now, the first-order change of the expectation value of any operator $A$ is $\delta\langle A\rangle
= i\int_0^{\delta t} dt\;\langle[H_{\rm int},A]\rangle_0=i\chi\,\delta t\langle[I_y,A]\rangle_0$, where $\langle\cdot\rangle_0$ denotes expectation values in the initial state. Thus
$\delta\langle O_x\rangle=i\chi\,\delta t\,\langle[I_y,O_x]\rangle_0$, and $\delta\langle I_x\rangle=i\chi\,\delta t\,\langle[I_y,I_x]\rangle_0$. Therefore the nuclear conversion factor can be written as
\begin{equation}
\chi_N=\frac{\langle[I_y,I_x]\rangle_0}{\langle[I_y,O_x]\rangle_0}\label{chiN}.
\end{equation}
As for the electron spin case, we can similarly write $\mathbf I|_F=d_F\,\mathbf F|_F$, where now $d_F = 1-c_F$, since $\mathbf{F}=\mathbf{S}+\mathbf{I}$. Thus, $d_a=2I/(2I+1)$ and $d_b=2(I+1)/(2I+1)$. For $^{87}$Rb it is $d_a=3/4$ and $d_b=5/4$. As before, neglecting hyperfine coherences we can write 
$\mathbf{I}\simeq d_a\mathbf{F}_a+d_b\mathbf{F}_b$, so that $\langle I_z\rangle\simeq d_a\langle F_{a,z}\rangle+d_b\langle F_{b,z}\rangle$. Using the commutation relations 
$[F_{\alpha,y},F_{\alpha,x}]=-iF_{\alpha,z}$ and $[F_{a},F_{b}]=0$, we obtain $[I_y,O_x]\simeq -i\left(d_a F_{a,z}-d_b F_{b,z}\right)$. Taking into account that $[I_y,I_x]=-iI_z$, and inserting these expressions into Eq.~\eqref{chiN} yields 
\begin{equation}
\chi_N
=\frac{d_a\langle F_{a,z}\rangle + d_b\langle F_{b,z}\rangle}
       {d_a\langle F_{a,z}\rangle - d_b\langle F_{b,z}\rangle}.
\label{kappaP}
\end{equation}

In the spin-temperature state $\rho\propto e^{\beta F_z}$, the electron spin polarization is 
$P = 2\langle S_z\rangle = \tanh(\beta/2)$, where $\beta$ is the inverse spin temperature~\,\cite{AppeltPRA1998}. Moreover, for $I=3/2$ it is $\langle F_{a,z}\rangle=P(3P^2+5)/2(P^2+1)$ and \mbox{$\langle F_{b,z}\rangle=P(1-P^2)/2(P^2+1)$}. Substituting these expressions into Eq.~\eqref{kappaP} we find
\begin{equation}
\chi_N(P) = \frac{2(P^2+5)}{7P^2+5}.
\label{eq:chiN}
\end{equation}
In contrast to the electronic case, the conversion factor $\chi_N$ is $P$-dependent.

In our experiment we optically pump $^{87}$Rb with $p_0=25~{\rm mW}$ of light resonant with the 780.2\,nm transition. The laser beam has an elliptical profile, with dimensions $7~{\rm mm}\times 3~{\rm mm}$, corresponding to a waist area
$A\approx16.5~{\rm mm}^2$. The optical pumping rate is $R=\int \sigma(\nu)\Phi(\nu)\,d\nu$, where $\Phi(\nu)$ is the photon flux spectral density of the pump light (number of photons per s per ${\rm m^2}$ per Hz of optical bandwidth).
The integrated absorption cross section is $\int\sigma(\nu)d\nu=\pi r_e c f_{\rm osc}$, where $r_e=2.82\times10^{-15}~{\rm m}$ is the classical electron radius and $f_{\rm osc}=0.696$ is the oscillator strength of the transition. The optical transition is pressure-broadened to a FWHM $\Gamma\approx11$\,GHz 
while the laser linewidth is $\approx1$\,MHz, so that the laser spectrum is much narrower than the atomic absorption profile. Approximating the broadened absorption profile by a Lorentzian lineshape, the on-resonance absorption cross section is $\sigma(\nu_0)= 2 r_e c f_{\rm osc}/\Gamma\approx 1.05\times 10^{-16}$\,m$^2$. For resonant optical pumping, the pumping rate may therefore be written as $R(l)=\big(p(l)/A h\nu_0\big)\,\sigma(\nu_0)$, where $\nu_0\approx3.84\times10^{14}$\,Hz is the transition frequency, and $p(l)=p_0e^{-\sigma(\nu_0)nl}$ is the optical power after a propagation length $l$ in the dense vapor, with $n=1.2\times10^{19}$/m$^3$ being the measured Rb density. 
The measured magnetic resonance linewidth is $\gamma = 2\pi\times4.5~{\rm kHz}$. Thus, the alkali-vapor spin polarization is $P(l)\approx R(l)/\big(R(l)+\gamma/2\big)$. 
Because the polarization varies across the overlap region of the pump and probe beams, we evaluate the mean conversion factor $\chi_N$ [Eq.\,(\ref{eq:chiN})] by averaging over the length of the interaction region, weighing with the polarization $P(l)$. We obtain $\chi_N = 1.99^{+0.01}_{-0.63}$. The uncertainty is dominated by imperfect knowledge of the optical depth $\sigma(\nu_0)\,n\,l$, which is primarily limited by the uncertainty in the density $n$. The lower bound ($1.37$) corresponds to a density that is three times smaller, whereas the upper bound ($2$) to a density three times larger. Note that $\chi_N \to 1$ as $P \to 1$; in our experiment, however, the mean polarization over the interaction region is small, so $\chi_N \to 2$.


Finally, the proton and neutron conversion factors pick up an additional factor $\sigma_p$ and $\sigma_n$, respectively, so that $\chi_p=\sigma_p\chi_N$, and $\chi_n=\sigma_n\chi_N$. To compute $\chi_p$ and $\chi_n$, we use results from semi-empirical nuclear models \cite{StadnikEPJC2015}, where it is found that the mean value of the proton and neutron spin in the nucleus of $^{87}$Rb is $\langle S_p\rangle\equiv\sigma_p I=0.376$ and $\langle S_n\rangle\equiv\sigma_n I=0.124$, respectively. Thus we obtain $\chi_p=0.499$ and $\chi_n=0.164$. Therefore our experiment can detect both ALP-proton and ALP-neutron couplings, albeit with about three times larger sensitivity for the proton.
\subsection{General case}
If two or all three couplings among the three possibilities (ALP-\textit{e}, ALP-\textit{p}, and ALP-\textit{n}) were to act simultaneously, it is clear that one measurement, e.g. the measurement of $O_x$ is not enough. To disentangle two or three couplings one could perform the measurement at different values of the spin-polarization $P$, in order to take advantage of the $P$-dependence of the nuclear conversion factor. Or one could change the probing altogether, e.g. use two or three probe-lasers and alter their wavelength to tune the couplings $g_a$ and $g_b$, so that one probes different linear combinations of $\mathbf{F}_a$ and $\mathbf{F}_b$. In any case, the present work considers a single APL-fermion coupling.

 \section{Constraints on ALP couplings\label{sec:constraints}}
An upper limit on the coupling $g_{\alpha ff}$ of Eq.\,(\ref{eq:pseudofield}) can be obtained using the determined limit on the magnetic field $\big| B_{\alpha_\perp}(\nu)\big|_{\rm lim}$: 
\begin{equation}
g_{\alpha ff}(\nu)=\frac{|\gamma_F|}{\chi_f \sqrt{2\hbar c \rho_{\mathrm{DM}}}}\frac{\big|{B}_{\alpha\perp}(\nu)\big]_{\rm lim}}{\mathbf{\vert v_{\alpha\perp}\vert}},
\label{eq:gConstraints}
\end{equation}

\noindent where $|\gamma_F|\approx 4.4\times10^{10} $\,rad\,s$^{-1}$T$^{-1}$ is the gyromagnetic ratio of the Rb ground hyperfine levels $F=1$ and $F=2$.
\noindent Within the two regions of spectral overlap in the  experimental sets (i.e.\,90$-$110\,kHz and 250$-$270\,kHz), a weighted value for $g_{aff}$ is computed to make optimal use of the overlapping data. 
The weight coefficients are the ones in the calculation of the merged experimental spectrum $P_n({\nu})$ of Eq.\,(\ref{eq:Ptapering}). We compute $g_{aff}$  in the first region as:

\begin{widetext}
\begin{align} 
g_{\alpha ff}(\nu)=\frac{|\gamma_F|}{\chi_f \sqrt{2\hbar c \rho_{\mathrm{DM}}}}\times %
\sqrt{w^{(1)}(\nu)\frac{\big[B_{\alpha\perp}^2(\nu)\big]_{\rm lim}^{(1)}}{\big[\mathbf{\vert v_{\alpha\perp}\vert}^2\big]^{(1)}}+w^{(2)}(\nu)\frac{\big[B_{\alpha\perp}^2(\nu)\big]^{(2)}_{\rm lim}}{\big[\mathbf{\vert v_{\alpha\perp}\vert}^2\big]^{(2)}}},
\label{eq:gConstraintsWeighted}
\end{align}
\end{widetext}

\noindent and likewise we obtain $g_{\alpha ff}$ in the second region [see Eq.\,(\ref{eq:Ptapering})]. 
 
 To evaluate $g_{aff}$, we require the projection $\vert \mathrm{v_{\alpha\perp}}\vert$ of the axion velocity on the  magnetometer's sensitive plane ($xy$). Because this projection is modulated (primarily by Earth's sidereal rotation), the value of $\vert \mathrm{v_{\alpha\perp}}\vert$ is matched in time to the data acquisition. The $\vert \mathrm{v_{\alpha\perp}}\vert$ is computed   with the Axionpy library \cite{LisantiPRD2021,AxionpyLibrary}, using the location of the laboratory (latitude of 35.3$^o$\,N, longitude  25.1$^o$E) together with the orientation of the magnetometer’s 
$z-$axis, which lies on the local horizontal plane pointing $\approx50^o$ east of north (see Fig.\,\ref{fig:apparatus} and Section\,\ref{sec:AlpProjection} in the Appendix). Over the course of the measurements used to produce the spectra of Fig.\,\ref{fig:B2spectrum}, the $\vert \mathrm{v_{\alpha\perp}}\vert$ varies between 47 and 
239 km/s. 

When the measurement time with a detector probing a given frequency $\nu_\alpha$ is smaller than the coherence time $\tau_a \approx \left( \nu_a v_a^2/{c^2} \right)^{-1}$, the stochastic fluctuations of the UDM field need to be taken into account when evaluating experimental sensitivity \cite{CentersNatComm2021, LisantiPRD2021}. 
We therefore consider whether our limits require an adjustment. Our magnetometer has a linewidth of $\approx4.5$\,kHz, which is far greater than the sub-Hz ALP linewidth. Given the 100-Hz step in the frequency scans and the respective 6-s dwell time per step, each frequency $\nu_{\alpha}$ is integrated effectively for time $T\approx270$\,s. This is much larger than the ALP coherence time $\tau_a$,  that varies between 2 and 17\,s across the search range. Consequently, the effects of stochasticity of the ALP field average out and are negligible in the present work.  

\begin{figure}[ht!]
    \centering    \includegraphics[width=\columnwidth]{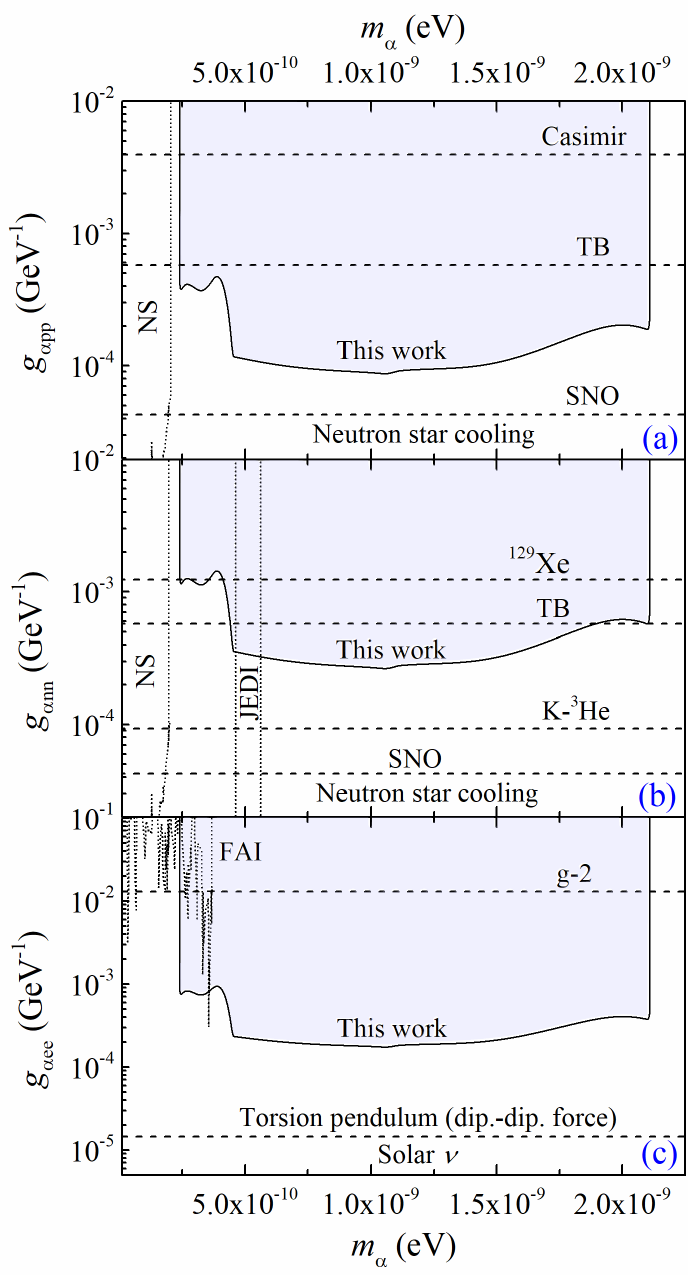}
 \caption{ 
 \small{ Exclusion plots for ALP--fermion couplings at the 95\% confidence level, with the region excluded by this work shown in color. 
(a) Limits on $g_{\alpha pp}$. For comparison, constraints from NASDUCK-SERF (NS)~\cite{BlochNatCom2023}; Casimir and torsion-balance (TB) experiments~\cite{MostepanenkoUniv2020,AdelbergerPRL2007}; solar-axion searches (SNO)~\cite{BhusalPRL2021}; and neutron-star cooling~\cite{BuschmannPRL2022} are shown. 
(b) Limits on $g_{\alpha nn}$. Previous constraints include K--$^{3}$He~\cite{VasilakisPRL2009}; $^{129}$Xe~\cite{SuPRL2024}; torsion-balance (TB) experiments~\cite{AdelbergerPRL2007}; NS~\cite{BlochNatCom2023}; JEDI~\cite{JEDI2022hxa}; and SNO~\cite{BhusalPRL2021}. 
(c) Limits on $g_{\alpha ee}$. Additional constraints shown are from the electron $g-2$ measurement~\cite{YanEPJC2019}; torsion-pendulum (dipole--dipole force) experiments~\cite{TerranoPRL2015}; a fermionic axion interferometer (FAI)~\cite{CresciniArxiv2023}; and solar-neutrino observations~\cite{GondoloPRD2009}. 
Data from previous works are taken from~\cite{Ohare2020axionlimits}.
}}
    \label{fig:gall}
\end{figure}

We show the derived limits on the proton, neutron and electron couplings in Fig.\,\ref{fig:gall}. They are calculated assuming ALP UDM makes up for all of the local DM density, and plotted over the searched-for mass range \mbox{$2.40\times10^{-10}\,\mathrm{eV}/c^{2}$--$2.11\times10^{-9}\,\mathrm{eV}/c^{2}$}.

For the coupling $g_{app}$, our result improves upon existing laboratory constraints from fifth-force searches over the whole  mass range investigated, with the obtained limit reaching its strongest value of $\approx 9\times10^{-5}$\,GeV$^{-1}$ at $m_{\alpha}\approx1.1\times10^{-9}$.
For the couplings $g_{ann}$ and $g_{aee}$, our constraints are not as stringent as those from fifth-force experiments, but they are complementary in scope. Fifth-force searches, such as torsion-balance and spin-based magnetometry experiments, are sensitive to static or quasi-static axion-mediated potentials and do not assume that axions constitute the local DM halo. By contrast, the present work probes the coupling of atomic spins to a coherently oscillating UDM ALP field. The two approaches therefore access the same underlying couplings through different physical observables.


While existing astrophysical bounds on ALP-fermion couplings are numerically stronger than the present results, they do not assume the presence of a local DM halo. Instead, they constrain ALP interactions through observations of astrophysical systems, independent of whether ALPs constitute the DM. Contrary to this, the present work provides a complementary laboratory constraint by directly probing the coupling of atomic spins to a coherently oscillating field in the DM halo, giving rise to a distinct signal at a frequency set by the ALP mass.




\section{Conclusions and outlook\label{sec:conclusions}}

We performed a laboratory search for ALP DM via its spin-dependent interaction with protons, neutrons and electrons, looking for a narrow-band, oscillatory magnetic-field signal in the radio-frequency range with an atomic magnetometer. We constrain the respective ALP couplings over the mass range \mbox{$2.40\times10^{-10}\,\mathrm{eV}/c^{2}$--$2.11\times10^{-9}\,\mathrm{eV}/c^{2}$}.


Our limits on the ALP-proton coupling improve on previous laboratory searches, for which the most sensitive constraints were obtained from fifth-force experiments. The bounds reported here on the ALP-neutron and ALP-electron couplings are less stringent, but complementary to existing fifth-force results, which probe ALP interactions without targeting the local DM halo. Although the present constraints are weaker than those derived from astrophysical observations, which likewise do not rely on a DM interpretation, our experiment directly targets ALPs as constituents of the Galactic DM halo.

The broadband frequency sensitivity of our radio-frequency atomic magnetometer enabled a wide search, reaching an ALP Compton frequency of 510 kHz, and the method can be extended to higher frequencies. In the present setup, sensitivity may be improved by increasing the effective magnetometric volume within the vapor cell (e.g., by using larger-area pump and probe beams) or by applying stronger optical pumping, which can induce light narrowing of the magnetic resonance and reduce its linewidth, thereby enhancing the signal-to-noise ratio.

Further sensitivity enhancement may be achieved by employing alternative operational modalities of the radio-frequency magnetometer, for example, by probing the axion-induced field using parametric-resonance techniques, which can suppress spin-exchange relaxation and yield a narrower effective linewidth \cite{XiaoPRL2024}.

\begin{acknowledgments}
    
We thank A. Zezas and G. Vasilakis for helpful discussions, and D. Stavrakaki for help with the project. This work was supported by the European Research Council (ERC) under the European Union Horizon 2020 research and innovation program (project YbFUN, grant agreement No 947696).\\

\end{acknowledgments}

\noindent\textbf{Author Contributions\\}
AR lead the apparatus development, and performed data taking and preliminary analysis; SN contributed to the apparatus development, and computed the ALP velocity data; IK calculated the atomic sensitivity to ALP couplings; DA analyzed data and supervised the work. All co-authors contributed to preparing the manuscript.\\

\clearpage

\appendix
\section{Apparatus\label{sec:Apparatus}}

\begin{figure*}[!t]
    \centering    \includegraphics[width=0.8\textwidth]{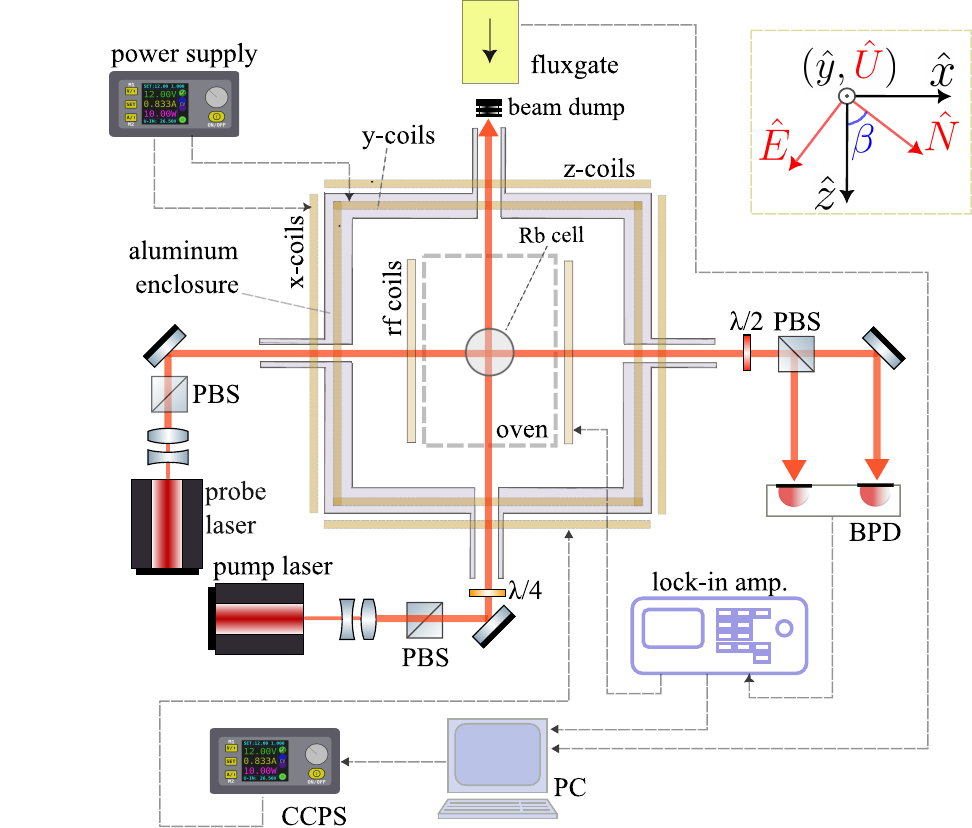}
 \caption{ 
 \small{ Experimental setup, showing a cross section of the aluminum enclosure housing the Rb vapor cell. The apparatus coordinate system (with the $xz$ plane lying on the local horizontal plane) is shown in relation to the local ENU system, whose $\hat{E}$,\,$\hat{N}$ axes are on the same plane. Abbreviations: PBS: polarizing beam splitter;$\lambda/2$: half-wave plate;  $\lambda/4$: quarter-wave plate; BPD: balanced photodetector; CPSS: computer-controlled power supply.}
}
    \label{fig:setup2D}
\end{figure*}

Fig.\,\ref{fig:setup2D} shows a schematic of the experimental apparatus. The Rb vapor cell is housed in a \(\approx 17\times 17\times 17\,\mathrm{cm}^3\) aluminum enclosure with a wall thickness of \(2\,\mathrm{cm}\). The cell is mounted in a oven operated at \(T=130\,^\circ\mathrm{C}\). Sets of approximately square coils mounted on the enclosure are used to null the ambient magnetic field along $\hat{x}$ and $\hat{y}$. A third coil set is used to apply a bias field along $\hat{z}$ to set the Larmor frequency, driven by a Thorlabs LDC210C current controller.

As discussed in Sec.~\ref{sec:experiment} of the main paper, active stabilization of the field along $\hat{z}$ is implemented in order to compensate for ambient field changes over time, so that the Larmor frequency during an experimental run can be accurately set. The stabilization involves measuring the field along $\hat{z}$ in the vicinity of the vapor cell (see~Fig.\,\ref{fig:setup2D}) with a fluxgate magnetometer (Stefan Mayer FLC 100) and applying feedback to the current driving the $\hat{z}$-set of coils, so that the reading of the fluxgate is set to a pre-determined value reflecting the sought Larmor frequency. The mapping of fluxgate readings to measured Larmor frequencies 
is done as follows: a) the magnetic resonance is recorded for fixed-current value by stepping the frequency of a test rf field around a nominal Larmor frequency, while the $\hat{z}$-directed field is measured with the fluxgate; b) the Larmor frequency is precisely determined from the spectrum, and the respective fluxgate reading is recorded; steps a) and b) are repeated for different z-current values; c) the  Larmor frequencies determined in b) for several current values are mapped to the respective values of the fluxgate readings.
This active compensation of the field along $\hat{z}$ ensures that the uncertainty in  setting the $\nu_L$ 
 value  during an hours-long UDM run is always smaller than 0.5\,kHz, or $\approx 10\%$ of the resonance linewidth. 


Approximately \(25\,\mathrm{mW}\) of light from a home-built external-cavity diode laser (ECDL) is used to polarize the \(^{87}\mathrm{Rb}\) atomic spins along $\hat{z}$. This laser is tuned in frequency to the center of the pressure-broadened \(780\)-nm transition. The pump beam is elliptical in profile with a \(7\times 3\)\,mm cross section. Probing of transverse magnetic fields on the $xy$-plane is performed using \(4\,\mathrm{mW}\) of light from a commercial ECDL (Toptica DL Pro), with a beam profile closely matched to the pump. The probe laser is detuned off resonance, about  \(\approx 100\,\mathrm{GHz}\) below the absorption peak. Faraday rotation is detected with a balanced polarimeter employing a Thorlabs PDB210A  photodetector. Its output is demodulated using a Zurich Instruments MFLI lock-in amplifier. The amplifier records the magnitude of the complex polarimeter signal \(|X+iY|\) with a $\tau=300$\,ms time constant, where \(X\) and \(Y\) are the in-phase and quadrature components, respectively.

\begin{figure}[ht!]
    \centering    \includegraphics[width=\columnwidth]{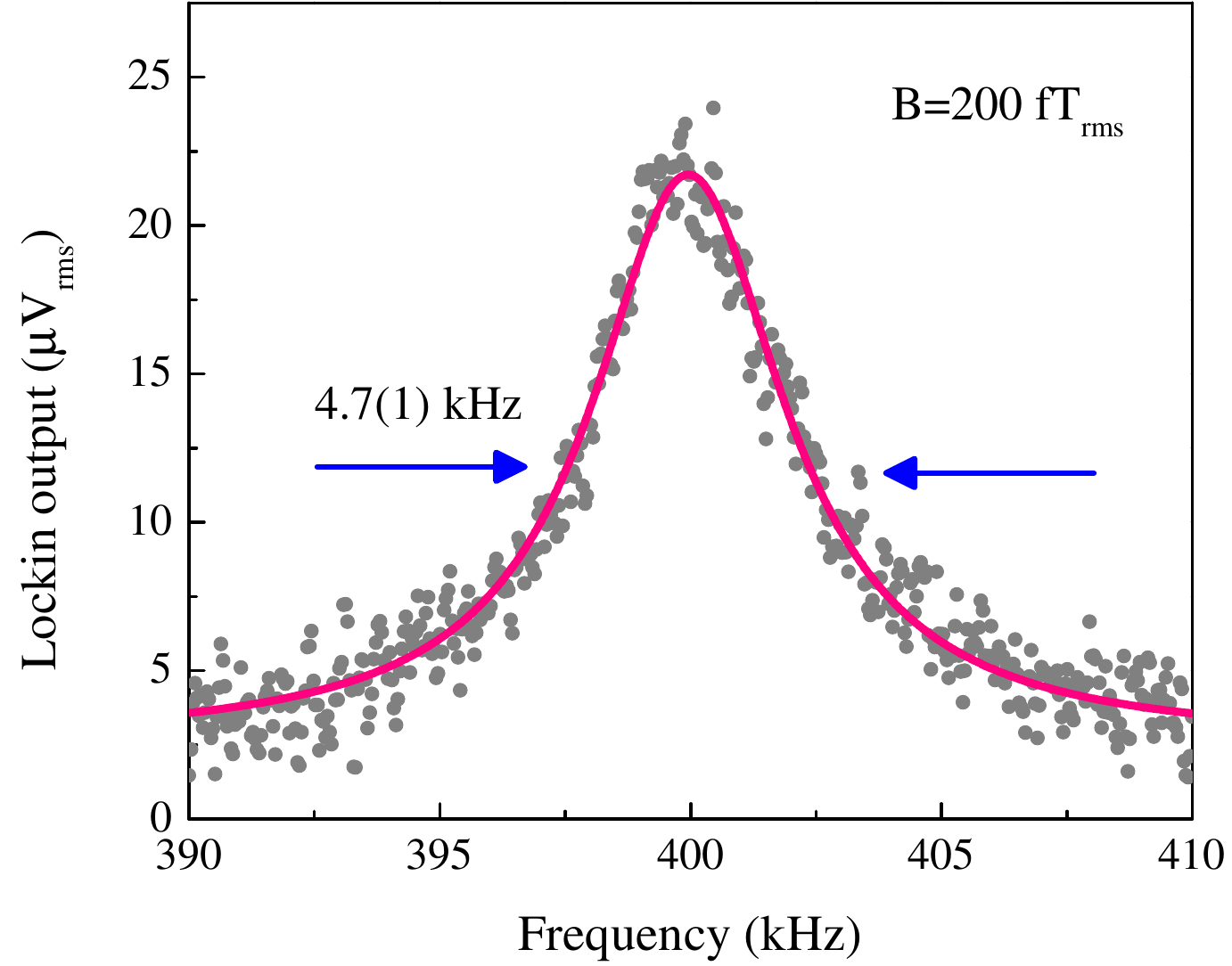}
 \caption{ 
 \small{Magnetic resonance recorded for a test field of 200--fT$_{rms}$ amplitude and 400--kHz Larmor frequency. The pink lines shows a Lorentzian fit to the data. }
}
    \label{fig:magres}
\end{figure}

Magnetometric response is characterized using calibrated rf coil pairs (\(\approx 10\times 6\,\mathrm{cm}^2\) each, separated by \(10\,\mathrm{cm}\)) that surround the oven assembly and generate rf fields along either the $\hat{x}$ or $\hat{y}$. Resonance spectra at a fixed Larmor frequency are acquired by stepping the frequency of the rf drive applied to one of the coil pairs via the LIA and recording the demodulated output. An example resonance is shown in Fig.~\ref{fig:magres}.

\section{Determination of the detection threshold $X_{thr}$\label{sec:DetThres}}

\begin{figure}[ht!]
    \centering    \includegraphics[width=\columnwidth]{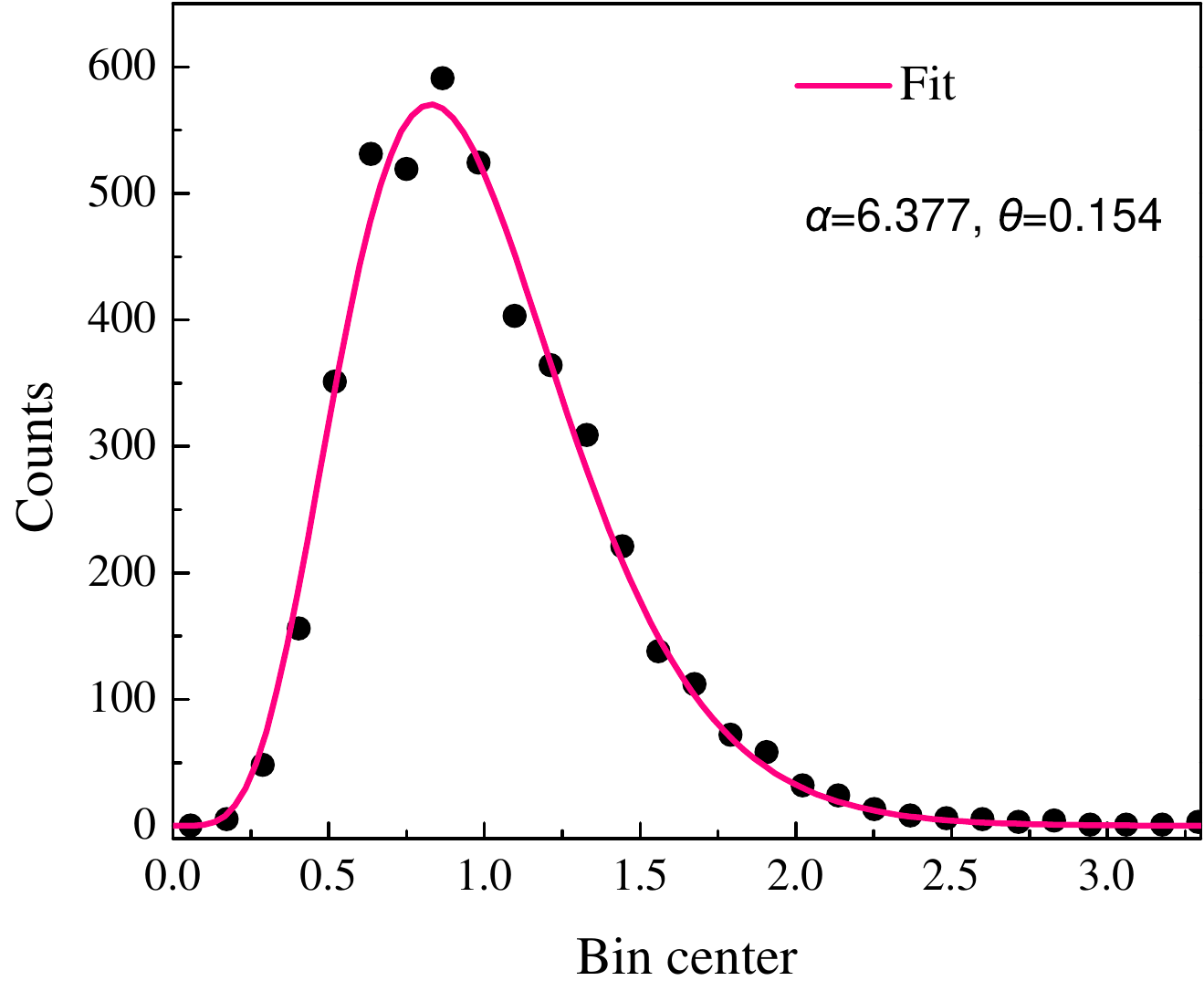}
 \caption{ 
 \small{ Frequency counts in the data of the normalized power spectrum $P_n(\nu)$ and fit with an unnormalized Gamma PDF with shape and scale parameters, $\alpha$ and $\theta$, respectively.}
}    \label{fig:Gammafit}
\end{figure}

The determined distribution function of the normalized experimental spectrum $P_n(\nu)$, is shown in Fig.\,\ref{fig:Gammafit}. It is well described by a Gamma distribution with the parameters indicated in the figure. This behavior is confirmed by simulations of the experimental signal acquisition and processing. For each frequency point, we average the simulated lock-in amplifier magnitude $\lvert X+iY\rvert$ over $K$ realizations, where $X$ and $Y$ are normally distributed variables. We then square this averaged magnitude and divide by the local mean noise level $N(\nu)$, following the same procedure used to construct $P_n(\nu)$. The simulations reproduce a Gamma distribution for values of $K$ comparable to those of the actual data acquisition, for which  $K$ is effectively $\approx 6$ (given the LIA time constant $\tau = 300$ ms, the 2$^{\rm nd}$-order filter applied to the demodulated signal, and the 4.5\,s of measurement time per frequency point). For $K\rightarrow\infty$, the simulated distribution becomes a Gaussian, as expected from the central-limit theorem.

We note that, while a $\chi^2$ distribution is expected for the squared magnitude $\lvert X+iY\rvert^2$ of normally distributed quadratures, our analysis involves averaging the magnitude prior to squaring. This nonlinear operation modifies the statistics, and for a finite number of averages $K$, the resulting distribution is well described by a Gamma distribution, as confirmed by our simulations. In the limit $K=1$, the standard $\chi^2$ behavior is recovered.

\begin{figure}[ht!]
    \centering    \includegraphics[width=\columnwidth]{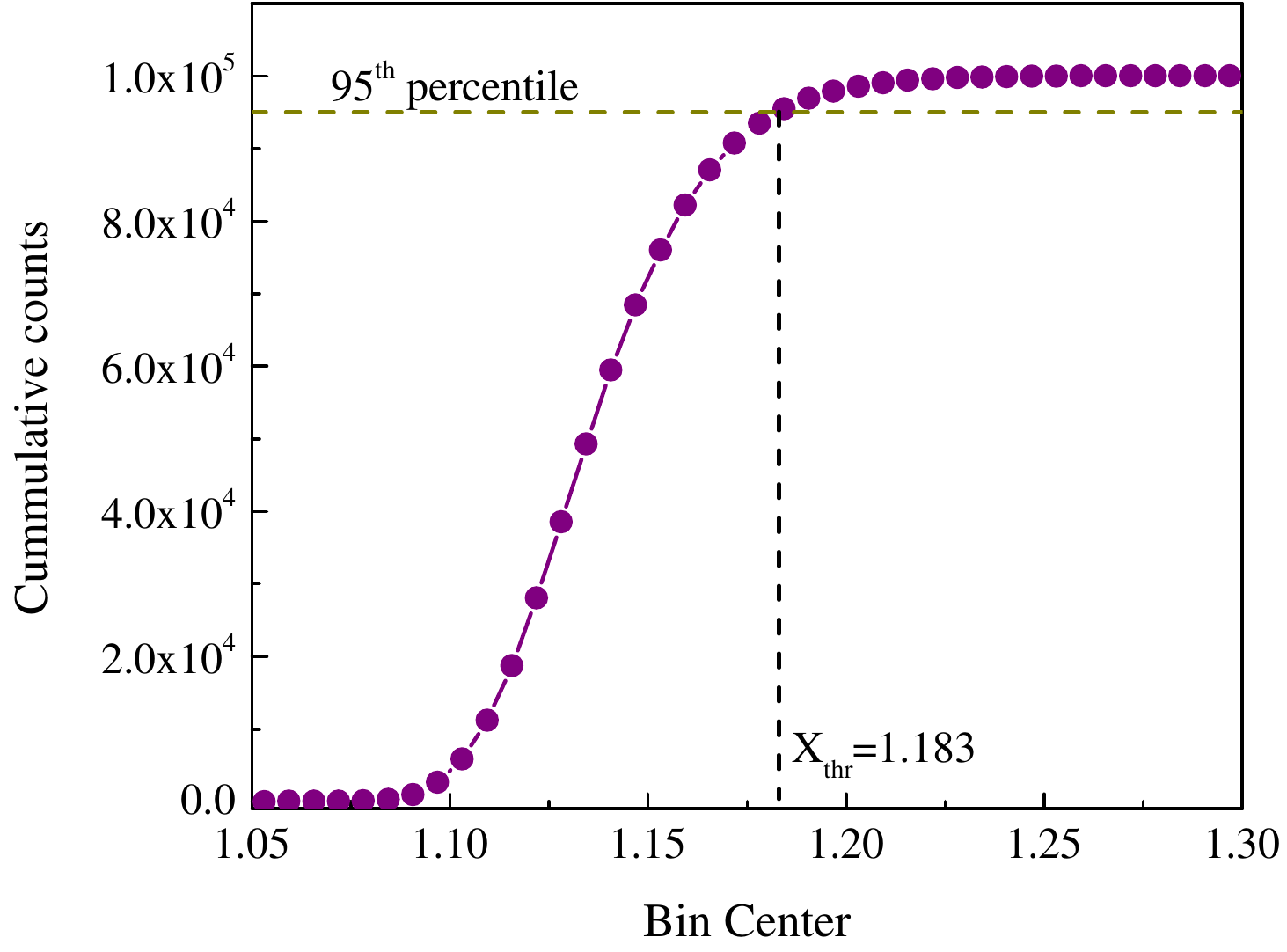}
 \caption{ 
 \small{Cumulative counts of the maximum value $X_{max}$ of simulated 10$^5$ $P_n(\nu)$ spectra, from which a detection threshold $X_{thr}$ in the experimental $P_n(\nu)$ spectrum is determined at the 95\% confidence level.}
}
    \label{fig:CDF}
\end{figure}

Simulated sets of $P_n(\nu)$ data, sampled from the empirically determined distribution of Fig.\,\ref{fig:Gammafit} are processed through the matched filter, and the cumulative distribution function of the maximum values  $X_{max}$ of the filtered spectra is used to define a detection threshold in the filtered experimental spectrum $P_n^f(\nu)$ of Fig.\,\ref{fig:FilteredB2}. This distribution is shown in Fig.\,\ref{fig:CDF}. From this, the threshold of $X_{\rm thr}=1.183$ is determined.

\section{Determination of $\vert v_{\alpha_\perp}\vert$\label{sec:AlpProjection}}
\begin{figure}[ht!]
    \centering    \includegraphics[width=\columnwidth]{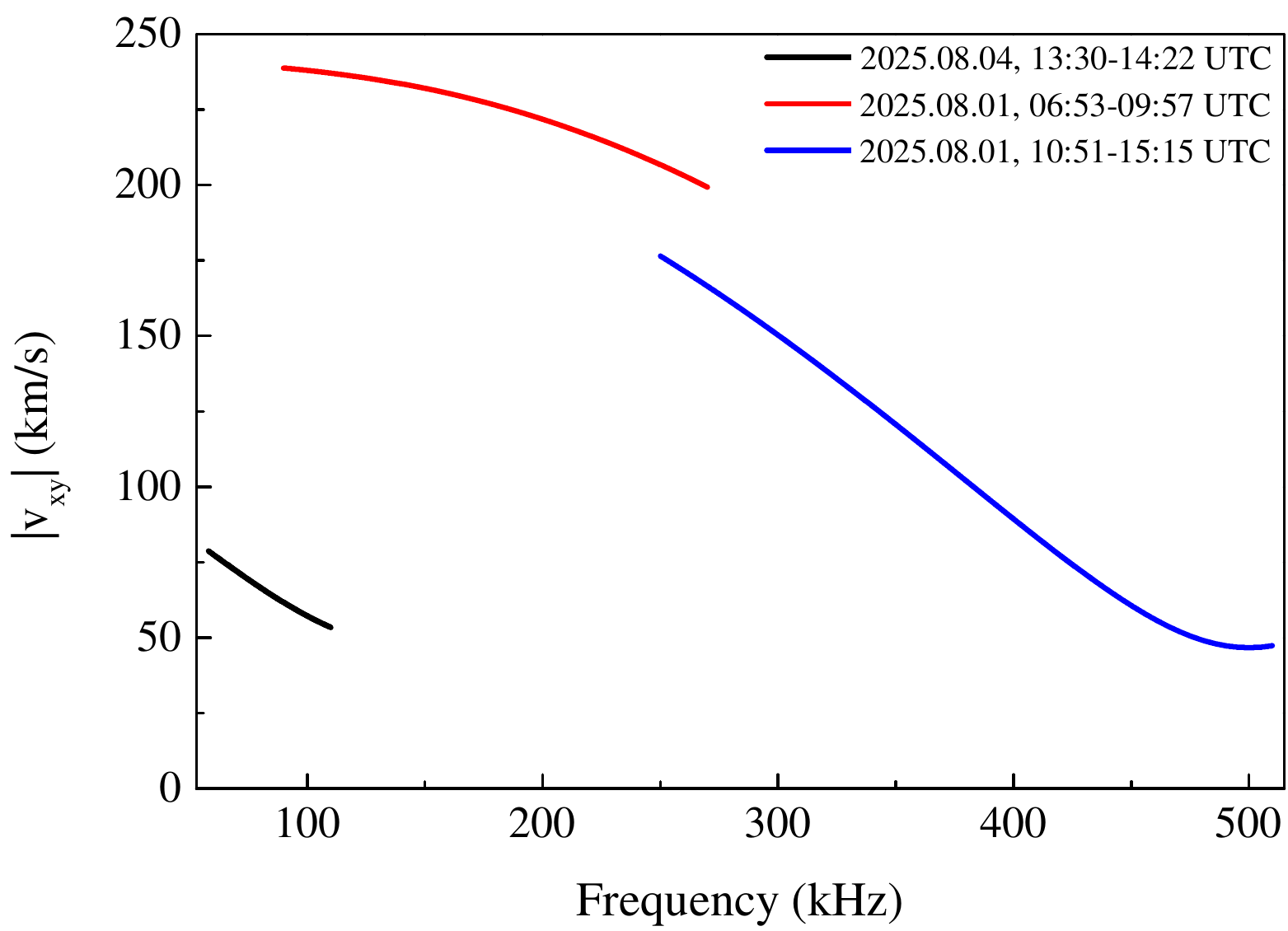}
 \caption{ 
 \small{Calculated ALP velocity projection on the magnetometer's sensitive $xy$ plane.  }
}
    \label{fig:vxy}
\end{figure}

We use the Axionpy library \cite{LisantiPRD2021,AxionpyLibrary} to compute the projection of the ALP velocity $\vert v_{\alpha_\perp}\vert$ on the apparatus sensitive $xy$ plane. The apparatus is located at latitude 35.3$^o$\,N and longitude  25.1$^o$E. The $xy$ plane is normal to the apparatus $\hat{z}$ axis, which lies on the local horizontal plane (Fig.\,\ref{fig:setup2D}). We relate the apparatus $xyz$ system to the local coordinate system with axes $\hat{N}$ (north), $\hat{E}$ (east), and $\hat{U}$ (local zenith), as shown in Fig.\,\ref{fig:setup2D}. The $\hat{z}$--axis points $\beta\approx 50^{o}$ east of north. The $\vert v_{\alpha_\perp}\vert$ is calculated as $\sqrt{\mathit{v_{\alpha x}^2}+\mathit{v_{\alpha y}^2}}$, where $\mathit{v_{\alpha x}}$ and $\mathit{v_{\alpha y}}$ are  the velocities on the $\hat{x}$ and $\hat{y}$ axis, respectively. Computations include the mean ALP speed  $\mathit{v_0}=220\,$\,km/s, the effect of the Earth's orbital speed revolving around the Sun at $\mathit{v_E}\approx30\,$\,km/s, and the effect due to the sidereal rotation of Earth,  which contributes by less than 1\,km/s. Calculations of $\vert v_{\alpha_\perp}\vert$ are shown in Fig.\,\ref{fig:vxy} for the respective data-acquisition intervals of the main runs presented in Fig.\,\ref{fig:B2spectrum}.

\section{Investigation of outlier peak\label{sec:Outliers}}

\begin{figure}[ht!]
    \centering    \includegraphics[width=\columnwidth]{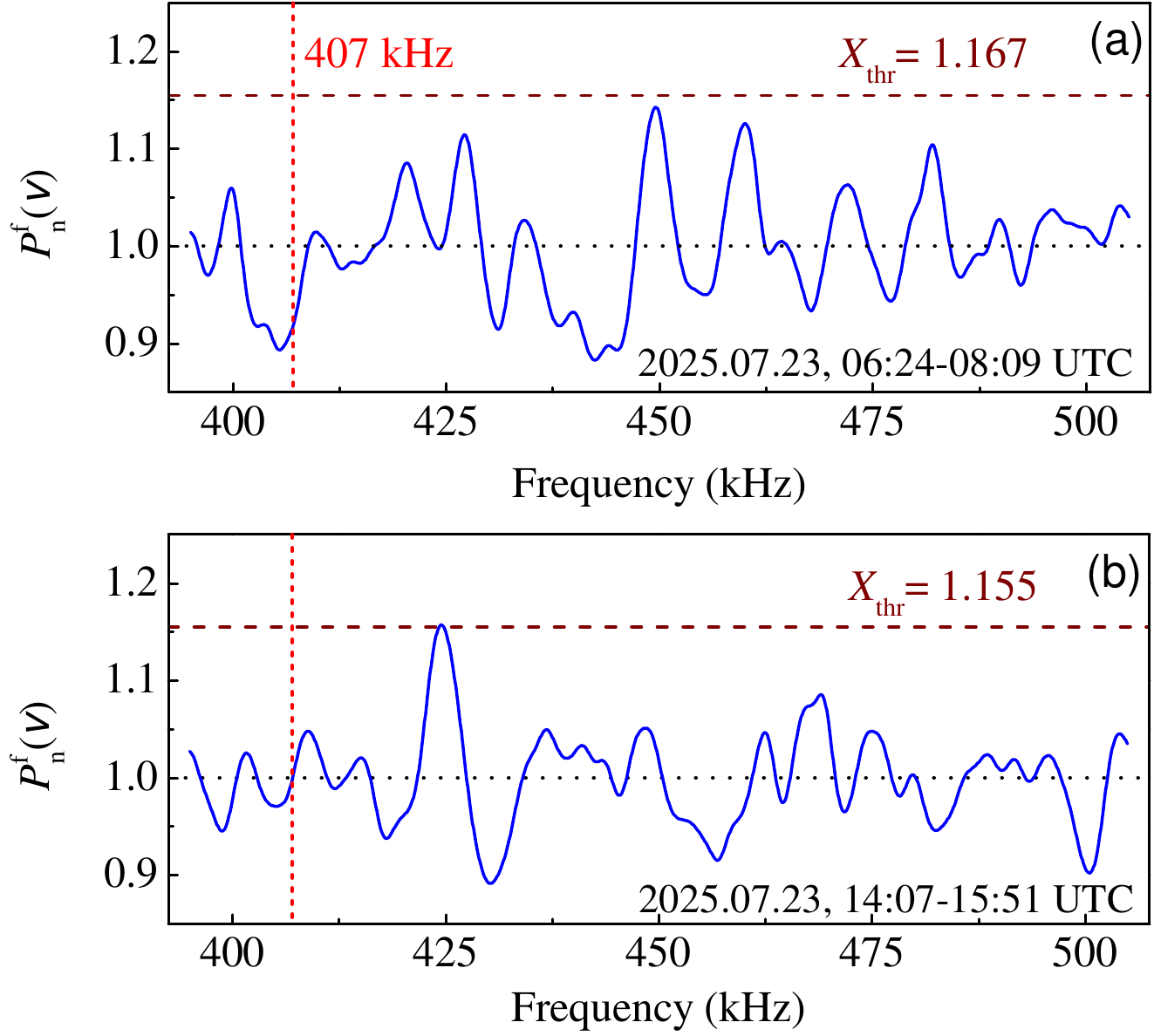}
 \caption{ 
 \small{Matched-filter output of the auxiliary datasets, with the corresponding detection thresholds at the 95\% CL. The vertical red line marks the outlier frequency from Fig.\,\ref{fig:FilteredB2}.
}
}
    \label{fig:outlierexclusion}
\end{figure}

We analyze additional magnetometer runs to investigate the outlier peak in the matched-filter output of the main experimental run, which exceeds the detection threshold (see Fig.\,\ref{fig:FilteredB2} in the main paper). Specifically, we consider two auxiliary datasets recorded over the frequency ranges (a) 395--505\,kHz and (b) 400--505\,kHz. Both datasets are processed with the same analysis procedure used for the primary runs presented in the main paper. The corresponding matched-filter outputs are shown in Fig.\,\ref{fig:outlierexclusion}, together with the detection threshold computed for each run. The outlier observed in Fig.\,\ref{fig:FilteredB2} at \(\approx 407\)\,kHz is not reproduced in either auxiliary dataset.

Constraints on the ALP-spin coupling are obtained from the ratio of the measured limit on a magnetic-field signal, \(\big[B_{\alpha}\big]_{\mathrm{lim}}\), to the projection \(|v_{\alpha\perp}|\) of the ALP velocity onto the magnetometer's sensitive \(xy\) plane (see Eq.~(\ref{eq:gConstraints}) in the main paper). Evaluating \(\big[B_{\alpha}\big]_{\mathrm{lim}}/|v_{\alpha\perp}|\) at the outlier frequency (407\,kHz), we obtain 0.13\,fT/(km/s) for the main experimental run, 0.07\,fT/(km/s) for run (a) in Fig.\,\ref{fig:outlierexclusion}, and 0.21\,fT/(km/s) for run (b). In particular, run (a) provides the most stringent constraint at this frequency.

\clearpage
\bibliographystyle{apsrev4-2}
\bibliography{bibliography.bib}

@PREAMBLE{
 "\providecommand{\noopsort}[1]{}" 
 # "\providecommand{\singleletter}[1]{#1}%" 
}

@book{KimballBook,
      title     = {The Search for Ultralight Bosonic Dark Matter},
      editor    = {Jackson Kimball, D. F. and van Bibber, K.},
      year      =  {2023},
      publisher = {Springer},
      address   = {New York}
    }

@article{GrahamARNPC2015,
    title = {{Experimental Searches for the Axion and Axion-Like Particles}},
    year = {2015},
    journal = {Annu. Rev. Nucl. Part. Sci.},
    author = {Graham, P. W. and Irastorza, I. G. and Lamoreaux, S. K. and Lindner, A. and Van Bibber, K. A.},
    pages = {485},
    volume = {65},
    doi = {10.1146/annurev-nucl-102014-022120}
}

@Article{BertoneNature2018,
author={Bertone, Gianfranco
and Tait, Tim M. P.},
title={A new era in the search for dark matter.},
journal={Nature},
year={2018},
month={Oct},
day={01},
volume={562},
number={7725},
pages={51-56},
issn={1476-4687},
doi={10.1038/s41586-018-0542-z},
url={https://doi.org/10.1038/s41586-018-0542-z}
}

@ARTICLE{Scargle1982,
       author = {{Scargle}, J.~D.},
        title = "{Studies in astronomical time series analysis. II. Statistical aspects of spectral analysis of unevenly spaced data.}",
      journal = {\apj},
         year = 1982,
        month = dec,
       volume = {263},
        pages = {835-853},
          doi = {10.1086/160554},
       adsurl = {https://ui.adsabs.harvard.edu/abs/1982ApJ...263..835S}
}

@article{CentersNatComm2021,
  year = {2021},
  doi = {10.1038/s41467-021-27632-7},
  month = dec,
  publisher = {Springer Science and Business Media {LLC}},
  volume = {12},
  number = {1},
  author = {P. Centers \textit{et al.}, G.},
  title = {Stochastic fluctuations of bosonic dark matter.},
pages={7321},
  journal = {Nat. Commun.}
}

@article{AntypasSnowmass2022,
  title={New Horizons: Scalar and Vector Ultralight Dark Matter.},
  author={Antypas \textit{et al.}, D},
  journal={arXiv:2203.14915},
  year={2022}
}

@article{PreskillPLB1983,
title = {Cosmology of the invisible axion.},
journal = {Phys. Lett. B},
volume = {120},
number = {1},
pages = {127-132},
year = {1983},
issn = {0370-2693},
doi = {https://doi.org/10.1016/0370-2693(83)90637-8},
url = {https://www.sciencedirect.com/science/article/pii/0370269383906378},
author = {J. Preskill and M. B. Wise and F. Wilczek}
}

@article{AbbottPLB1983,
title = {A cosmological bound on the invisible axion.},
journal = {Phys. Lett. B},
volume = {120},
number = {1},
pages = {133-136},
year = {1983},
issn = {0370-2693},
doi = {https://doi.org/10.1016/0370-2693(83)90638-X},
url = {https://www.sciencedirect.com/science/article/pii/037026938390638X},
author = {L.F. Abbott and P. Sikivie}
}

@article{DinePLB1983,
title = {The not-so-harmless axion.},
journal = {Phys. Lett. B},
volume = {120},
number = {1},
pages = {137-141},
year = {1983},
issn = {0370-2693},
doi = {https://doi.org/10.1016/0370-2693(83)90639-1},
url = {https://www.sciencedirect.com/science/article/pii/0370269383906391},
author = {Michael Dine and Willy Fischler}
}

@book{Demtroder,
  author    = {W. Demtröder}, 
  title     = {Laser Spectroscopy},
  publisher = {Springer},
  year      = 2015,
  volume    = 2,
  address   = {Berlin-Heidelberg},
  edition   = {5th}
  }

@article{Bloch:2021vnn,
    author = "Bloch, I. M. and Ronen, G. and Shaham, R. and Katz, O. and Volansky, T. and Katz, O.",
    collaboration = "NASDUCK",
    title = "{New constraints on axion-like dark matter using a Floquet quantum detector.}",
    eprint = "2105.04603",
    archivePrefix = "arXiv",
    primaryClass = "hep-ph",
    doi = "10.1126/sciadv.abl8919",
    journal = "Sci. Adv.",
    volume = "8",
    number = "5",
    pages = "abl8919",
    year = "2022"
}

@article{BuschmannPRL2022,
  title = {Upper Limit on the QCD Axion Mass from Isolated Neutron Star Cooling},
  author = {Buschmann, Malte and Dessert, Christopher and Foster, Joshua W. and Long, Andrew J. and Safdi, Benjamin R.},
  journal = {Phys. Rev. Lett.},
  volume = {128},
  issue = {9},
  pages = {091102},
  numpages = {9},
  year = {2022},
  month = {Mar},
  publisher = {American Physical Society},
  doi = {10.1103/PhysRevLett.128.091102},
  url = {https://link.aps.org/doi/10.1103/PhysRevLett.128.091102}
}

@article{KimballNJP2015,
doi = {10.1088/1367-2630/17/7/073008},
url = {https://dx.doi.org/10.1088/1367-2630/17/7/073008},
year = {2015},
month = {jul},
publisher = {IOP Publishing},
volume = {17},
number = {7},
pages = {073008},
author = {Kimball, D F Jackson},
title = {Nuclear spin content and constraints on exotic spin-dependent couplings},
journal = {New J. Phys.}
}

@article{LeePRX2023,
  title = {Laboratory Constraints on the Neutron-Spin Coupling of feV-Scale Axions},
  author = {Lee, Junyi and Lisanti, Mariangela and Terrano, William A. and Romalis, Michael},
  journal = {Phys. Rev. X},
  volume = {13},
  issue = {1},
  pages = {011050},
  numpages = {25},
  year = {2023},
  month = {Mar},
  publisher = {American Physical Society},
  doi = {10.1103/PhysRevX.13.011050},
  url = {https://link.aps.org/doi/10.1103/PhysRevX.13.011050}
}

@Article{GMNatComm2025,
author={Gavilan-Martin \textit{et al.}, D.},
title={Searching for dark matter with a spin-based interferometer},
journal={Nat. Commun.},
year={2025},
month={May},
day={28},
volume={16},
number={1},
pages={4953},
issn={2041-1723},
doi={10.1038/s41467-025-60178-6},
url={https://doi.org/10.1038/s41467-025-60178-6}
}

@article{CresciniArxiv2023,
  author       = {Crescini, Nicol\`o},
  title        = {The Fermionic Axion Interferometer},
  journal      = {arXiv preprint},
  volume       = {arXiv:2311.16364},
  year         = {2023},
  note         = {[hep-ex]},
  url          = {https://arxiv.org/abs/2311.16364}
}

@article{SavukovPRL2005,
  title = {Tunable Atomic Magnetometer for Detection of Radio-Frequency Magnetic Fields},
  author = {Savukov, I. M. and Seltzer, S. J. and Romalis, M. V. and Sauer, K. L.},
  journal = {Phys. Rev. Lett.},
  volume = {95},
  issue = {6},
  pages = {063004},
  numpages = {4},
  year = {2005},
  month = {Aug},
  publisher = {American Physical Society},
  doi = {10.1103/PhysRevLett.95.063004},
  url = {https://link.aps.org/doi/10.1103/PhysRevLett.95.063004}
}

@Article{BlochNatCom2023,
author={Bloch, Itay M.
and Shaham, Roy
and Hochberg, Yonit
and Kuflik, Eric
and Volansky, Tomer
and Katz, Or},
title={Constraints on axion-like dark matter from a SERF comagnetometer},
journal={Nat. Commun.},
year={2023},
month={Sep},
day={18},
volume={14},
number={1},
pages={5784},
issn={2041-1723},
doi={10.1038/s41467-023-41162-4},
url={https://doi.org/10.1038/s41467-023-41162-4}
}

@article{KimballPRD2016,
  title = {Magnetic shielding and exotic spin-dependent interactions},
  author = {Jackson Kimball, D. F. and Dudley, J. and Li, Y. and Thulasi, S. and Pustelny, S. and Budker, D. and Zolotorev, M.},
  journal = {Phys. Rev. D},
  volume = {94},
  issue = {8},
  pages = {082005},
  numpages = {8},
  year = {2016},
  month = {Oct},
  publisher = {American Physical Society},
  doi = {10.1103/PhysRevD.94.082005},
  url = {https://link.aps.org/doi/10.1103/PhysRevD.94.082005}
}

@article{TerranoPRL2015,
  title = {Short-Range, Spin-Dependent Interactions of Electrons: A Probe for Exotic Pseudo-Goldstone Bosons},
  author = {Terrano, W. A. and Adelberger, E. G. and Lee, J. G. and Heckel, B. R.},
  journal = {Phys. Rev. Lett.},
  volume = {115},
  issue = {20},
  pages = {201801},
  numpages = {5},
  year = {2015},
  month = {Nov},
  publisher = {American Physical Society},
  doi = {10.1103/PhysRevLett.115.201801},
  url = {https://link.aps.org/doi/10.1103/PhysRevLett.115.201801}
}

@article{GondoloPRD2009,
  title = {Solar neutrino limit on axions and keV-mass bosons},
  author = {Gondolo, Paolo and Raffelt, Georg G.},
  journal = {Phys. Rev. D},
  volume = {79},
  issue = {10},
  pages = {107301},
  numpages = {4},
  year = {2009},
  month = {May},
  publisher = {American Physical Society},
  doi = {10.1103/PhysRevD.79.107301},
  url = {https://link.aps.org/doi/10.1103/PhysRevD.79.107301}
}

@article{NavasPRD2024,
    author = "Navas, S. and others",
    collaboration = "Particle Data Group",
    title = "{Review of particle physics}",
    doi = "10.1103/PhysRevD.110.030001",
    journal = "Phys. Rev. D",
    volume = "110",
    number = "3",
    pages = "030001",
    year = "2024"
}

@article{MarshPhysRep2016,
title = {},
journal = {Phys. Rep.},
volume = {643},
pages = {1-79},
year = {2016},
note = {},
issn = {0370-1573},
doi = {https://doi.org/10.1016/j.physrep.2016.06.005},
url = {https://www.sciencedirect.com/science/article/pii/S0370157316301557},
author = {David J.E. Marsh}
}

@article{LisantiPRD2021,
  title = {Stochastic properties of ultralight scalar field gradients},
  author = {Lisanti, Mariangela and Moschella, Matthew and Terrano, William},
  journal = {Phys. Rev. D},
  volume = {104},
  issue = {5},
  pages = {055037},
  numpages = {15},
  year = {2021},
  month = {Sep},
  publisher = {American Physical Society},
  doi = {10.1103/PhysRevD.104.055037},
  url = {https://link.aps.org/doi/10.1103/PhysRevD.104.055037}
}

@article{WUPRL2019,
  title = {Search for Axionlike Dark Matter with a Liquid-State Nuclear Spin Comagnetometer},
  author = {Wu \textit{et al.}, T.},
  journal = {Phys. Rev. Lett.},
  volume = {122},
  issue = {19},
  pages = {191302},
  numpages = {6},
  year = {2019},
  month = {May},
  publisher = {American Physical Society},
  doi = {10.1103/PhysRevLett.122.191302},
  url = {https://link.aps.org/doi/10.1103/PhysRevLett.122.191302}
}

@article{WeiRPP2025,
  author       = {Wei \textit{et al.}, K.},
  title        = {Dark matter search with a resonantly-coupled hybrid spin system},
  journal      = {Rep. Prog. Phys.},
  volume       = {88},
  number       = {5},
  year         = {2025},
  doi          = {10.1088/1361-6633/adca52},
}

@article{BudkerPRX2013,
  title = {Proposal for a Cosmic Axion Spin Precession Experiment (CASPEr)},
  author = {Budker, Dmitry and Graham, Peter W. and Ledbetter, Micah and Rajendran, Surjeet and Sushkov, Alexander O.},
  journal = {Phys. Rev. X},
  volume = {4},
  issue = {2},
  pages = {021030},
  numpages = {10},
  year = {2014},
  month = {May},
  publisher = {American Physical Society},
  doi = {10.1103/PhysRevX.4.021030},
  url = {https://link.aps.org/doi/10.1103/PhysRevX.4.021030}
}

@article{WalterPRD2025,
  title = {Search for axionlike dark matter using liquid-state nuclear magnetic resonance},
  author = {Walter \textit{et al.}, J.},
  journal = {Phys. Rev. D},
  volume = {112},
  issue = {5},
  pages = {052008},
  numpages = {15},
  year = {2025},
  month = {Sep},
  publisher = {American Physical Society},
  doi = {10.1103/39nc-vr9m},
  url = {https://link.aps.org/doi/10.1103/39nc-vr9m}
}

@article{WechslerAnnRevAstro2018,
   author = "Wechsler, Risa H. and Tinker, Jeremy L.",
   title = "The Connection Between Galaxies and Their Dark Matter Halos", 
   journal= "Annu. Rev. Astron. Astrophys.",
   year = "2018",
   volume = "56",
   number = "Volume 56, 2018",
   pages = "435-487",
   doi = "https://doi.org/10.1146/annurev-astro-081817-051756",
   url = "https://www.annualreviews.org/content/journals/10.1146/annurev-astro-081817-051756",
   publisher = "Annual Reviews",
   issn = "1545-4282",
   type = "Journal Article",
  }

@article{FosterPRD2018,
  title = {Revealing the dark matter halo with axion direct detection},
  author = {Foster, Joshua W. and Rodd, Nicholas L. and Safdi, Benjamin R.},
  journal = {Phys. Rev. D},
  volume = {97},
  issue = {12},
  pages = {123006},
  numpages = {34},
  year = {2018},
  month = {Jun},
  publisher = {American Physical Society},
  doi = {10.1103/PhysRevD.97.123006},
  url = {https://link.aps.org/doi/10.1103/PhysRevD.97.123006}
}

@Article{StadnikEPJC2015,
author={Stadnik, Y. V.
and Flambaum, V. V.},
title={Nuclear spin-dependent interactions: searches for WIMP, axion and topological defect dark matter, and tests of fundamental symmetries},
journal={Eur. Phys. J. C.},
year={2015},
month={Mar},
day={07},
volume={75},
number={3},
pages={110},
issn={1434-6052},
doi={10.1140/epjc/s10052-015-3326-8},
url={https://doi.org/10.1140/epjc/s10052-015-3326-8}
}

@article{
GarconSciAdv2019,
author = {Garcon \textit{et al.}, A. },
title = {Constraints on bosonic dark matter from ultralow-field nuclear magnetic resonance},
journal = {Sci. Adv.},
volume = {5},
number = {10},
pages = {eaax4539},
year = {2019},
doi = {10.1126/sciadv.aax4539},
URL = {},
eprint = {}}

@article{BrubakerPRD2018,
  title = {HAYSTAC axion search analysis procedure},
  author = {Brubaker, B. M. and Zhong, L. and Lamoreaux, S. K. and Lehnert, K. W. and van Bibber, K. A.},
  journal = {Phys. Rev. D},
  volume = {96},
  issue = {12},
  pages = {123008},
  numpages = {34},
  year = {2017},
  month = {Dec},
  publisher = {American Physical Society},
  doi = {10.1103/PhysRevD.96.123008},
  url = {https://link.aps.org/doi/10.1103/PhysRevD.96.123008}
}

@misc{AxionpyLibrary,
  author       = {{M.T. Moschella \textit{et al.}}},
  title        = {Axionpy},
  howpublished = {\url{https://github.com/mtmoschella/axionpy/tree/main/axionpy}},
  note         = {},
  year         = {}
}

@article{MouloudakisPRA2023,
  title = {Interspecies spin-noise correlations in hot atomic vapors},
  author = {Mouloudakis \textit{et al.}, K.},
  journal = {Phys. Rev. A},
  volume = {108},
  issue = {5},
  pages = {052822},
  numpages = {22},
  year = {2023},
  month = {Nov},
  publisher = {American Physical Society},
  doi = {10.1103/PhysRevA.108.052822},
  url = {https://link.aps.org/doi/10.1103/PhysRevA.108.052822}
}

@article{AppeltPRA1998,
  title = {Theory of spin-exchange optical pumping of ${}^{3}\mathrm{He}$ and ${}^{129}\mathrm{Xe}$},
  author = {Appelt, S. and Baranga, A. Ben-Amar and Erickson, C. J. and Romalis, M. V. and Young, A. R. and Happer, W.},
  journal = {Phys. Rev. A},
  volume = {58},
  issue = {2},
  pages = {1412--1439},
  numpages = {0},
  year = {1998},
  month = {Aug},
  publisher = {American Physical Society},
  doi = {10.1103/PhysRevA.58.1412},
  url = {https://link.aps.org/doi/10.1103/PhysRevA.58.1412}
}

@article{AdelbergerPRL2007,
  title = {Particle-Physics Implications of a Recent Test of the Gravitational Inverse-Square Law},
  author = {Adelberger, E. G. and Heckel, B. R. and Hoedl, S. and Hoyle, C. D. and Kapner, D. J. and Upadhye, A.},
  journal = {Phys. Rev. Lett.},
  volume = {98},
  issue = {13},
  pages = {131104},
  numpages = {4},
  year = {2007},
  month = {Mar},
  publisher = {American Physical Society},
  doi = {10.1103/PhysRevLett.98.131104},
  url = {https://link.aps.org/doi/10.1103/PhysRevLett.98.131104}
}

@Article{MostepanenkoUniv2020,
AUTHOR = {Mostepanenko, Vladimir M. and Klimchitskaya, Galina L.},
TITLE = {The State of the Art in Constraining Axion-to-Nucleon Coupling and Non-Newtonian Gravity from Laboratory Experiments},
JOURNAL = {Universe},
VOLUME = {6},
YEAR = {2020},
NUMBER = {9},
ARTICLE-NUMBER = {147},
URL = {https://www.mdpi.com/2218-1997/6/9/147},
ISSN = {2218-1997},
DOI = {10.3390/universe6090147}
}

@article{BhusalPRL2021,
  title = {Searching for Solar Axions Using Data from the Sudbury Neutrino Observatory},
  author = {Bhusal, Aagaman and Houston, Nick and Li, Tianjun},
  journal = {Phys. Rev. Lett.},
  volume = {126},
  issue = {9},
  pages = {091601},
  numpages = {7},
  year = {2021},
  month = {Mar},
  publisher = {American Physical Society},
  doi = {10.1103/PhysRevLett.126.091601},
  url = {https://link.aps.org/doi/10.1103/PhysRevLett.126.091601}
}

@misc{Ohare2020axionlimits,
  author    = {O'Hare, Ciaran},
  title     = {{cajohare/AxionLimits: AxionLimits} (Version v1.0)},
  year      = {2020},
  publisher = {Zenodo},
  doi       = {10.5281/zenodo.3932430},
  url       = {https://doi.org/10.5281/zenodo.3932430}
}

@Article{YanEPJC2019,
author={Yan, H.
and Sun, G. A.
and Peng, S. M.
and Guo, H.
and Liu, B. Q.
and Peng, M.
and Zheng, H.},
title={Constraining exotic spin dependent interactions of muons and electrons},
journal={Eur. Phys. J. C.},
year={2019},
month={Nov},
day={25},
volume={79},
number={11},
pages={971},
issn={1434-6052},
doi={10.1140/epjc/s10052-019-7442-8},
url={https://doi.org/10.1140/epjc/s10052-019-7442-8}
}

@article{VasilakisPRL2009,
  title = {Limits on New Long Range Nuclear Spin-Dependent Forces Set with a $\mathbf{K}\mathrm{\text{\ensuremath{-}}}^{3}\mathrm{He}$ Comagnetometer},
  author = {Vasilakis, G. and Brown, J. M. and Kornack, T. W. and Romalis, M. V.},
  journal = {Phys. Rev. Lett.},
  volume = {103},
  issue = {26},
  pages = {261801},
  numpages = {4},
  year = {2009},
  month = {Dec},
  publisher = {American Physical Society},
  doi = {10.1103/PhysRevLett.103.261801},
  url = {https://link.aps.org/doi/10.1103/PhysRevLett.103.261801}
}

@article{SuPRL2024,
  title = {New Constraints on Axion-Mediated Spin Interactions Using Magnetic Amplification},
  author = {Su, Haowen and Jiang, Min and Wang, Yuanhong and Huang, Ying and Kang, Xiang and Ji, Wei and Peng, Xinhua and Budker, Dmitry},
  journal = {Phys. Rev. Lett.},
  volume = {133},
  issue = {19},
  pages = {191801},
  numpages = {8},
  year = {2024},
  month = {Nov},
  publisher = {American Physical Society},
  doi = {10.1103/PhysRevLett.133.191801},
  url = {https://link.aps.org/doi/10.1103/PhysRevLett.133.191801}
}

@article{JEDI2022hxa,
  title = {First Search for Axionlike Particles in a Storage Ring Using a Polarized Deuteron Beam},
  author = {Karanth \textit{et al.}, S.},
  collaboration = {JEDI Collaboration},
  journal = {Phys. Rev. X},
  volume = {13},
  issue = {3},
  pages = {031004},
  numpages = {28},
  year = {2023},
  month = {Jul},
  publisher = {American Physical Society},
  doi = {10.1103/PhysRevX.13.031004},
  url = {https://link.aps.org/doi/10.1103/PhysRevX.13.031004}
}

@article{IrastorzaSD2018,
title = {New experimental approaches in the search for axion-like particles},
journal = {Prog. Part. Nucl. Phys.},
volume = {102},
pages = {89-159},
year = {2018},
issn = {0146-6410},
doi = {https://doi.org/10.1016/j.ppnp.2018.05.003},
url = {https://www.sciencedirect.com/science/article/pii/S014664101830036X},
author = {Igor G. Irastorza and Javier Redondo}}

@article{SavukovJMR2007,
title = {Detection of NMR signals with a radio-frequency atomic magnetometer},
journal = {J. Magn. Reson.},
volume = {185},
number = {2},
pages = {214-220},
year = {2007},
issn = {1090-7807},
doi = {https://doi.org/10.1016/j.jmr.2006.12.012},
url = {https://www.sciencedirect.com/science/article/pii/S1090780706004101},
author = {I.M. Savukov and S.J. Seltzer and M.V. Romalis}
}

@article{XiaoPRL2024,
  title = {Radio-Frequency Magnetometry Based on Parametric Resonances},
  author = {Xiao, Wei and Liu, Xiyu and Wu, Teng and Peng, Xiang and Guo, Hong},
  journal = {Phys. Rev. Lett.},
  volume = {133},
  issue = {9},
  pages = {093201},
  numpages = {6},
  year = {2024},
  month = {Aug},
  publisher = {American Physical Society},
  doi = {10.1103/PhysRevLett.133.093201},
  url = {https://link.aps.org/doi/10.1103/PhysRevLett.133.093201}
}

\end{document}